\documentclass[12pt,aps,pra,eqsecnum,psfig,amssymb,subeqn,graphicx]{article}
\usepackage{amssymb}
\def\be{\begin{equation}}
\def\ee{\end{equation}}
\def\beqa{\begin{eqnarray}}
\def\eeqa{\end{eqnarray}}
\def\wph{{\omega}}

\begin{document}
\baselineskip 0.76cm

\title{Invariant vector fields and the prolongation method for supersymmetric quantum systems}
\author{Nibaldo Alvarez M.\thanks{email address: alvarez@dms.umontreal.ca} and V\'eronique Hussin\thanks{email address: hussin@dms.umontreal.ca} \\
\small{ Centre de Recherches Math\'ematiques et d\'epartement de
Math\'ematiques et de statistique, }  \\ \small{Universit\'e  de
Montr\'eal, C.P. 6128, Succ.~Centre-ville, Montr\'eal (Qu\'ebec),
H3C 3J7, Canada} } \maketitle

\begin{abstract}
The kinematical and dynamical symmetries of equations describing the time evolution of quantum systems
like the supersymmetric harmonic oscillator in one space dimension and the interaction of a non-relativistic
spin one-half particle in a constant magnetic field are reviewed from the point of view of the vector field prolongation
method. Generators of supersymmetries are then introduced so that we get Lie superalgebras of symmetries and
supersymmetries. This approach does not require the introduction of Grassmann valued differential equations
but a specific matrix realization and the concept of dynamical symmetry.
The Jaynes-Cummings model and supersymmetric generalizations are then studied.
We show how it is closely related to the preceding models. Lie algebras of symmetries and supersymmetries are
also obtained.

\end{abstract}


\newpage

\section{Introduction} The symmetries of a system of ordinary differential equations
(ODEÕs) or partial differential equations (PDEÕs) are usually
obtained by using the so-called prolongation method of vector
fields \cite{kn:olver,kn:bluman}. It consists of finding the
infinitesimal generators which close the maximal invariant Lie
algebra of the system of equations. The corresponding symmetry
group is the Lie group of local transformations of independent and
dependent variables which leaves invariant the system under
consideration. Such a system may be associated with the wave
equation of some quantum model. The independent variables are the
usual space-time coordinates while the dependent ones are the
components of the wave function. The symmetries may be related to
the so-called kinematical Lie algebra \cite{kn:niederer} of the
quantum system.

Now, if we have in mind supersymmetric (SUSY) quantum models \cite{kn:witten,kn:CrRi}, the question is how
to find them from this prolongation method.  We answer this question
by considering first some standard examples where the kinematical Lie superalgebras are known.
This is the case of the SUSY harmonic oscillator
in one space dimension (see \cite{kn:durand,kn:BDH} and reference therein) and the Pauli equation,
in two space dimensions, describing the motion of a
non-relativistic spin one-half particle in a constant magnetic field \cite{kn:BDH}.
An important part of this work is concerned by the study of symmetries of the Jaynes-Cummings (JC)
model \cite{kn:JaCu} based on the same approach. Let us recall that the JC model, which consists of an
idealized description of the interaction of a quantized electromagnetic field and an atomic system
with two levels, is closely related with the two models considered before. An interesting point is that
it can be made SUSY in a non-trivial manner and our approach will clarify this point and will make
the connection with different works on this subject \cite{kn:AnLe,kn:BuRa}.

At the classical level, Grassmann-valued differential equations have
been introduced \cite{kn:MaRa, kn:mathieu,kn:RoKe,kn:BrDa} and the prolongation method has been extended
to include Grassmann independent and dependent variables \cite{kn:AHW,kn:AAH}.
For example, SUSY extensions of Korteweg-de-Vries and other equations have been studied
and maximal invariant Lie superalgebras have been obtained.

At the quantum level, the problem is somewhat different. The SUSY system is nothing
but a set of PDEÕs with the usual independent and dependent variables. So it is really
of the type where the usual prolongation method can be used and the vector fields obtained
close a Lie algebra. The non-trivial question we ask is how to get the generators which are associated
with supersymmetries from this method and which, together with the symmetry generators, close a
Lie superalgebra.

To clarify the context we are working with, let us here recall the prolongation method
\cite{kn:olver,kn:bluman} for determining the symmetries of a system of $m$ PDEÕs of order $n$ of the type
\be\label{gsyseq}
\Delta^{(k)} [x ;
u_\alpha, {u_\alpha}^{(1)}_{ x_{j_1 }}, {u_\alpha }^{(2)}_{
x_{j_1} x_{j_2 }} , \cdots, {u_\alpha}^{(n)}_{ x_{j_1} x_{j_2}
\cdots x_{j_{n}  } } ] =0, \qquad k=1,2, \ldots ,m,
\ee
with $p$ independent variables $x_j \ (j=1,2,\ldots p),$
and $q$ dependent variables $u_\alpha(x) \ ( \alpha=1,2, \ldots  q).$
The derivatives of the dependent variables are defined as
 \be
{u_\alpha }^{(l)}_{ x_{j_1} x_{j_2} \cdots x_{j_l } } \equiv
u_\alpha^{(l)} \equiv {\partial^l {u_\alpha } (x) \over
\partial x_{j_1} \partial  x_{j_2} \cdots \partial x_{j_l}} ,
\qquad   1 \le l \le n,
\ee
where the integers  $ j_r \ (r=1,2, \ldots , l)$
are such that $ 0 \le j_r \le p. $

The Lie group of local transformations of independent and dependent variables which
leave invariant such a system is obtained by performing the following infinitesimal
transformation on the independent and dependent variables:
\beqa
\tilde x_j = x_j +
\epsilon
\sum_j \xi_j (x, u_\alpha )  + {\cal O}(\epsilon^2 ), \label{tildex} \\
{\tilde u}_\alpha (\tilde x , {\tilde u}_\beta ) = u_\alpha (x,
u_\beta ) + \epsilon \phi_\alpha (x , u_\beta ) + {\cal O}(\epsilon^2 ), \label{tildeu}
\eeqa
Assuming that they satisfy, at first order in $\epsilon$, the equation
\be\label{gsyseqtilde} \Delta^{(k)}
[\tilde x ; {\tilde u}_\alpha, {{\tilde u}_{\alpha_{{\tilde
x}_{j_1 }}}^{(1)}, {{\tilde u}_{\alpha_{{\tilde
x}_{j_1 } {\tilde x}_{j_2}}}^{(2)}, \cdots, {{\tilde u}_{\alpha_{{\tilde
x}_{j_1 }{\tilde x}_{j_2} \cdots {\tilde x}_{j_{n}}}^{(n)}}}}}] =0,
 \ee
for $k=1,2, \ldots ,m $ and when the $u_\alpha (x)$ solve
the system (\ref{gsyseq}), we can find the functions $\xi_j$ and $\phi_\alpha$.
A practical way to do it is to introduce the vector field
\be \label{vectfield}
v = \sum_{j =1}^p \xi_j (x, u_\beta ) \partial_{x_j}  + \sum_{\alpha
=1}^q \phi_\alpha (x, u_\beta ) \partial_{u_\alpha}   ,
\ee
associated with the transformations (\ref{tildex}) and (\ref{tildeu}) and
define the $n$th order prolongation of $v$ as
\be
pr^{(n)} v = v + \sum_{\alpha =1}^q \sum_{J_l, \,
l=1,2,\ldots,n} \phi_\alpha^{J_l} (x, u_\beta , u^{(1)}_\beta ,
u^{(2)}_\beta, \ldots, u^{(n)}_\beta )
 \partial_{u_\alpha^{J_l}} , \ee
where $ J_l=  (x_{j_1}, x_{j_2}, \ldots , x_{j_l}) $ is the multi-index notation
for the differentiation with respect to the $x_{j}$ and
$\partial_{u_\alpha^{J_l}} \equiv  \partial_{u^{(l)}_\alpha}$.
Note that the
coefficients $\phi_\alpha^{J_l}$ satisfy the following recurrence relation
\be \label{relrec}
\phi_\alpha^{J_l,x_k} =
D_{x_k} \phi_\alpha^{J_l} - \sum_{j=1}^p (D_{x_k} \xi_j )
{\partial u_\alpha^{J_l} \over
\partial x_j},
\ee where $ D_{x_k}$ is the total derivative with respect to
$x_k$. The infinitesimal criterion for invariance
(\ref{gsyseqtilde}) may then be written as \be \label{applpro}
pr^{(n)} v \{\Delta^{(k)} [x ; u_\alpha, {u_\alpha}^{(1)}_{ x_{j_1
}}, {u_\alpha }^{(2)}_{ x_{j_1} x_{j_2 }} , \cdots,
{u_\alpha}^{(n)}_{ x_{j_1} x_{j_2} \cdots x_{j_{n}  } } ] \}=0,
\qquad k=1,2, \ldots ,m, \ee when the $u_\alpha (x)$ satisfy
(\ref{gsyseq}). The condition (\ref{applpro}) gives a set of PDE's
called the determining equations which can be solved to get the
explicit form of the functions $\xi_j$ and $\phi_\alpha$ in
(\ref{vectfield}). The resolution may lead to different
possibilities: no nontrivial solutions, a finite number of
integration constants or that the general solution depends on
arbitrary functions. Let us also mention that we have the
following properties of the vector field prolongations: \be
pr^{(n)} (c_1 v_1 + c_2 v_2 + \cdots + c_n v_m ) =pr^{(n)} c_1 v_1
+ pr^{(n)} c_2 v_2 + \cdots + pr^{(n)} c_n v_m \ee and \be
pr^{(n)}[v_1 ,v_2 ] = [pr^{(n)}v_1 , pr^{(n)} v_2 ]. \ee

The contents of the paper is thus described as follows. Section 2 is devoted to the construction of
invariant vector fields for the SUSY harmonic oscillator in one dimension. It admits a large set of symmetries
and the integration of vector fields gives a matrix realization of the symmetry generators which is essential
in order to find the generators of supersymmetries. The corresponding kinematical and dynamical invariance
superalgebras will be recovered in this context. In Section 3, the model of a
non-relativistic spin-${1\over 2}$ particle in a constant magnetic field is studied. It can be reduced
to a two dimensional model  and shows a similar behaviour than the SUSY harmonic oscillator. The symmetry algebra
and superalgebra are obtained from the prolongation of vector fields method and connected to the preceding case. In Section 4,
we start with a quantum evolution equation which is a realization of the JC model and determine the
invariant vector fields and the associated invariant algebra. The connection with the preceding models is very helpful to get a Lie
superalgebra of symmetries for a generalized JC model. In section 5, we propose a SUSY version of this model
and give the corresponding symmetries and supersymmetries. We also make the connection with preceding attemps to get SUSY
JC models.

\section{The SUSY harmonic oscillator}
\label{sec-uno} The first set of equations we are considering is
the one associated with the SUSY harmonic oscillator in one space
dimension. The corresponding Schr\"{o}dinger evolution equation is
\be \label{schopersusy} \left( i \partial_t -  H_{{\rm SUSY}}
\right ) \Psi (t,x) = 0. \ee Let us mention that along this work
we use the convention that $\hbar=1$. The SUSY Hamiltonian
\cite{kn:JwAsSaVh} is given by: \be H_{{\rm SUSY}} = \biggl( - {1
\over 2 M} {\partial^2 \over
\partial x^2 } + {1 \over 2} M \omega^2 x^2 \biggr) \sigma_0  - {\omega \over 2 } \sigma_3,
\ee
where $\sigma_0 $ is the identity matrix and $\sigma_3 = \pmatrix{ 1 &
  0 \cr 0 & -1 \cr}.$ The wave function takes the form
\be \label{wavefunction} \Psi (t,x)= \pmatrix{\psi_1 (t,x) \cr
\psi_2 (t,x)}, \quad \psi_1, \psi_2 \in L^2 ({\mathbb R}). \ee It
is convenient to write the equation (\ref{schopersusy}) as a set
of two equations \be \label{schrosusyuno}
 i {\partial \psi_\alpha  \over \partial t }   +
{1 \over 2 M} {\partial^2 \psi_\alpha \over \partial x^2 } - {1
\over 2} M \omega^2 x^2  \psi_\alpha + {\omega_\alpha \over 2 }
\psi_\alpha = 0, \qquad \alpha =1,2, \ee where we have set
$\omega_1 =\omega $ et $\omega_2=- \omega$.

The kinematical and dynamical symmetries and supersymmetries have
been largely studied
\cite{kn:niederer,kn:CrRi,kn:durand,kn:BDH,kn:wybourne} but these
approaches were different than the one we want to apply. Indeed for
the usual harmonic oscillator, Niederer \cite {kn:niederer} has
first shown that the maximal kinematical algebra is the semi-direct
sum $ so(2,1)  \dotplus h(2),$  where $h(2)$ is the usual
Heisenberg-Weyl algebra. The maximal dynamical algebra
\cite{kn:wybourne}, defined as the one associated with the
degeneracy group of the model, is given by \ensuremath{
sp(2){\dotplus } h(2)} and includes the preceding kinematical
algebra. The dynamical and kinematical superalgebras of the SUSY
version coincide in this one-dimensional case and is given by
\ensuremath{osp(2/2){ \dotplus } sh(2/2)} \cite {kn:durand,kn:BDH}.
We will show how to recover these structures starting from the
prolongation method of vector fields applied to the system (\ref
{schrosusyuno}).

\subsection{Prolongation method and invariant vector fields}
A standard way of applying the prolongation method to a system containing complex valued functions is to express
the components of the wave function (\ref{wavefunction}) as
 \be \label{chanvar}
\psi_1 (t,x) = u_1 (t,x) e^{i \nu_1 (t,x)},
 \quad \psi_2 (t,x) = u_2 (t,x) e^{i \nu_2 (t,x)},
\ee
where $ u_1,\ u_2, \ \nu_1 $ and $\nu_2 $ are real functions of $t$ and $x$.
Inserting (\ref{chanvar}) into (\ref{schrosusyuno}) and separating the real and complex parts of the resulting equations, we
are led to a set of four coupled equations in $ u_1,\ u_2, \ \nu_1 $ and $\nu_2 $.
The vector field (\ref{vectfield}) may be written explicitly as
\be \label{cha-vec-susy}
v= \xi_1 \partial_t + \xi_2 \partial_x + \phi_1 \partial_{u_1} + \phi_2 \partial_{u_2} +  \varphi_1
\partial_{\nu_1} + \varphi_2 \partial_{\nu_2},
\ee where $ \xi_j \ (j=1,2),\  \phi_\alpha  $ and $\varphi_\alpha
\ (\alpha =1,2) $ are real functions which depend on $t,\ x,\ u_1
,\ u_2 ,\ \nu_1 $ and $\nu_2 $.

A simpler way of solving the
problem is to consider the set (\ref{schrosusyuno}) together with
its complex conjugated \be \label{barschrosusyuno} - i {\partial
{\bar \psi}_\alpha  \over
\partial t }   +  {1 \over 2 M} {\partial^2 {\bar \psi}_\alpha
\over \partial x^2 } - {1 \over 2} M \omega^2 x^2 {\bar
\psi_\alpha} + {\omega_\alpha \over 2 } {\bar \psi_\alpha}  = 0,
\qquad \alpha=1,2. \ee Now the corresponding vector field takes
the form
 \be
\label{barchampvec} v= \xi_1 \partial_t + \xi_2  \partial_x +
\Phi_1 \partial_{\psi_1}  + {\bar \Phi}_1 \partial_{\bar \psi_1} +
\Phi_2 \partial_{\psi_2}  + {\bar \Phi}_2 \partial_{\bar \psi_2},
\ee where now $ \xi_j \ (j=1,2)$ are real functions of the
variables $t,\ x,\ \psi_1, \ \psi_2, \ {\bar \psi}_1 $ and $ {\bar
\psi}_2$ and $ \Phi_\alpha, \ {\bar \Phi}_\alpha \ (\alpha =1,2)$
are possible complex valued functions of these variables.  In
terms of these variables, the second order prolongation of $v$
takes the form: \be pr^{(2)} v = v + \sum_{\alpha=1}^2 \left(
\Phi^t_\alpha
\partial_{\psi_{\alpha, t}} + \Phi^x_\alpha \partial_{\psi_{\alpha, x}} +
\Phi^{tt}_\alpha \partial_{\psi_{\alpha,t t}} + \Phi^{tx}_\alpha
\partial_{\psi_{\alpha,t x}} + \Phi^{xx}_\alpha \partial_{\psi_{\alpha,xx}} \right) +
\left[ {\rm c.c} \right] , \ee where, for example, $\psi_{\alpha,
t}$ is the usual partial derivative of  $\psi_{\alpha}$ with
respect to $t$. Applying this prolongation to the system
consisting in equations (\ref{schrosusyuno}) and
(\ref{barschrosusyuno}), we get \beqa \label{phiuno} i
\Phi^t_\alpha + {1\over 2M} \Phi^{xx}_\alpha - {1 \over 2M}
\omega^2 x^2 \Phi_\alpha  +
{\omega_\alpha \over 2 } \Phi_\alpha - M \omega^2 x \xi_1 \psi_\alpha &=&0,  \\
 \label{barphiuno}
- i {\bar \Phi}^t_\alpha + {1\over 2M} {\bar \Phi}^{xx}_\alpha -
{1 \over 2M}  \omega^2 x^2 {\bar \Phi}_\alpha  + {\omega_\alpha
\over 2 } {\bar \Phi}_\alpha - M \omega^2 x \xi_1 {\bar
\psi}_\alpha & =& 0, \eeqa where we have \be \Phi^t_\alpha = D_t
\Phi_\alpha - \left( D_t \xi_1 \right ) \psi_{\alpha,t}  - \left(
D_t  \xi_2 \right ) \psi_{\alpha, x}, \ee \beqa \Phi^{xx}_\alpha =
(D_x \Phi^x_\alpha) - (D_x  \xi_1 ) \psi_{\alpha,xt}- (D_x  \xi_2
) \psi_{\alpha,xx}, \eeqa with \be \Phi^x_\alpha = D_x \Phi_\alpha
- \left(  D_x \xi_1 \right ) \psi_{\alpha,t}  - \left(  D_x \xi_2
\right ) \psi_{\alpha, x}, \ee together with their complex
conjugated and for $\alpha=1,2$. Inserting these expressions into
the system (\ref{phiuno}-\ref{barphiuno}), taking into account the
equations (\ref{schrosusyuno}) and  (\ref{barschrosusyuno}) and
identifying to zero the coefficients of the partial derivatives,
we get a set of determining equations which will give the
functions $ \xi_j , \Phi_\alpha $ and $ {\bar  \Phi}_\alpha. $
Solving these equations, we get: \beqa
\xi_1 (t) &=& {1 \over 2 \omega} (\delta_1 \sin 2\omega t - \delta_2 \cos 2\omega t ) + \delta_3, \\
\xi_2 (t,x) &=& {1 \over 2} (\delta_1 \cos 2\omega t + \delta_2
\sin 2\omega t ) x + \delta_4 \cos \omega t + \delta_5 \sin \omega
t, \eeqa which are effectively real functions depending only on
the coordinates $t$ and $x$. We also have \beqa
\Phi_1 (t ,x, \psi_1 , \psi_2 ) &=&  A_0 (t,x) + A_1(t , x)  \psi_1 + A_2 (t)  \psi_2,  \\
\Phi_2 (t ,x, \psi_1 , \psi_2 ) &=&  B_0 (t,x) + B_1(t )  \psi_1 +
B_2 (t,x)  \psi_2, \eeqa where \beqa \nonumber A_1 (t,x ) &=& - {1
\over 4} \left(e^{-2i \omega t} + 2i M \omega x^2 \sin 2\omega t
\right ) \delta_1 -
{i \over 4} \left( e^{-2i \omega t}- 2 M \omega x^2 \cos 2\omega t  \right ) \delta_2  \\
&-& i M \omega x (\delta_4 \sin \omega t - \delta_5 \cos \omega t ) + \delta_{13} + i \delta_6 , \\
A_2(t) & = & (\delta_7 - i\delta_{10}  )  e^{i \omega t}   , \\
B_1 (t) &=& ( \delta_8 - i \delta_{11} )  e^{-i\omega t }, \qquad \\
\nonumber B_2 (t,x ) &=&  - {1 \over 4} \left(e^{2i \omega t} + 2i
M \omega x^2 \sin 2\omega t  \right ) \delta_1 +
{i \over 4} \left( e^{2i \omega t}+ 2 M \omega x^2 \cos 2\omega t  \right ) \delta_2  \\
&-& i M \omega x (\delta_4 \sin \omega t - \delta_5 \cos \omega t
) + \delta_9 + i \delta_{12}. \eeqa The parameters $ \delta_i \ (
i=1,2, \ldots, 13) $ are all real and the functions ${\bar
\Phi}_1,\ {\bar \Phi}_2$ are in fact the complex conjugate of
${\Phi}_1,\ {\Phi}_2$. The functions $ A_0 (t,x) , \ B_0 (t,x)$
and their conjugated ${\bar A}_0 (t,x),\ {\bar B}_0 (t,x) $ are
such that they satisfy respectively (\ref{schrosusyuno}) and
(\ref{barschrosusyuno}) for $\psi_1=A_0$ and $\psi_2=B_0$.

The infinitesimal generators of the invariance finite dimensional
Lie algebra are thus easily obtained using the preceding equations
and (\ref{barchampvec}). We get \beqa \tilde X_1 &=& {1 \over 2
\omega } \sin 2\omega t \partial_t  + {x \over 2}  \cos 2\omega t
\partial_x - {1 \over 4}  \cos 2\omega t
( \psi_1 \partial_{\psi_1} + {\bar \psi}_1 \partial_{\bar \psi_1} + \psi_2 \partial_{\psi_2} + {\bar \psi}_2 \partial_{\bar \psi_2} )  \nonumber \\
&-& i {M \omega x^2 \over 2} \sin 2\omega t \left ( (   \psi_1
\partial_{\psi_1} - {\bar \psi}_1 \partial_{\bar \psi_1} )
+  ( \psi_2 \partial_{\psi_2}  - {\bar \psi}_2 \partial_{\bar \psi_2} ) \right) \nonumber \\
&+&    {i\over 4} \sin 2\omega t    \left( ( \psi_1
\partial_{\psi_1} - {\bar \psi}_1 \partial_{\bar \psi_1})
-  (  \psi_2 \partial_{\psi_2} - {\bar \psi}_2 \partial_{\bar \psi_2}) \right ), \nonumber \\
\tilde X_2 &=& -{ 1 \over 2 \omega } \cos 2\omega t  \partial_t +
{x \over 2}  \sin 2\omega t \partial_x - {1 \over 4}  \sin 2\omega
t
( \psi_1 \partial_{\psi_1} + {\bar \psi}_1 \partial_{\bar \psi_1} + \psi_2 \partial_{\psi_2} + {\bar \psi}_2 \partial_{\bar \psi_2} )  \nonumber \\
&+&  i{M \omega x^2 \over 2} \cos 2\omega t \left (  (   \psi_1
\partial_{\psi_1} - {\bar \psi}_1 \partial_{\bar \psi_1} )
 +  ( \psi_2 \partial_{\psi_2}  - {\bar \psi}_2 \partial_{\bar \psi_2} ) \right) \nonumber \\
&-&    {i\over 4} \cos 2\omega t    \left(  ( \psi_1
\partial_{\psi_1} - {\bar \psi}_1 \partial_{\bar \psi_1})
 -  (  \psi_2 \partial_{\psi_2} - {\bar \psi}_2 \partial_{\bar \psi_2}) \right ), \nonumber \\
\tilde X_3 &=& \partial_t,\nonumber\\
\tilde X_4 &=& \cos \omega t \partial_x - i M \omega x \sin \omega
t \left( ( \psi_1 \partial_{\psi_1} - {\bar \psi}_1
\partial_{\bar \psi_1})
+ (  \psi_2 \partial_{\psi_2} - {\bar \psi}_2 \partial_{\bar \psi_2}) \right ) ,\nonumber \\
\tilde X_5 &=& \sin \omega t \partial_x +  i M \omega x \cos
\omega t \left( ( \psi_1 \partial_{\psi_1} - {\bar \psi}_1
\partial_{\bar \psi_1})
 +  (  \psi_2 \partial_{\psi_2} - {\bar \psi}_2 \partial_{\bar \psi_2}) \right ) , \nonumber \\
\tilde X_6 &=& i    ( \psi_1 \partial_{\psi_1} - {\bar \psi}_1 \partial_{\bar \psi_1}), \nonumber \\
\tilde X_7 &=&  e^{i\omega t} \psi_2 \partial_{\psi_1} + e^{-i\omega t}   {\bar \psi}_2 \partial_{\bar \psi_1}, \nonumber \\
\tilde X_8 &=&  e^{-i\omega t} \psi_1 \partial_{\psi_2} + e^{i\omega t}   {\bar \psi}_1 \partial_{\bar \psi_2}, \nonumber \\
\tilde X_9 &=& \psi_2 \partial_{\psi_2}  + {\bar \psi}_2 \partial_{\bar \psi_2},\nonumber \\
\tilde X_{10} &=& i \left( e^{-i\omega t}   {\bar \psi_2} \partial_{\bar \psi_1} - e^{i\omega t} \psi_2 \partial_{\psi_1}   \right),\nonumber \\
\tilde X_{11} &=& i \left( e^{i\omega t}   {\bar \psi_1} \partial_{\bar \psi_2} - e^{-i\omega t} \psi_1 \partial_{\psi_2}  \right), \nonumber \\
\tilde X_{12} &=& i ( \psi_2 \partial_{\psi_2}  - {\bar \psi_2} \partial_{\bar \psi_2} ), \nonumber\\
\tilde X_{13} &=& ( \psi_1 \partial_{\psi_1} + {\bar \psi}_1 \partial_{\bar \psi_1}).\nonumber
\eeqa

If we come back to the real variables $ u_\alpha $ et $ \nu_\alpha\ (\alpha=1,2) $ introduced in (\ref{chanvar}),
we have the following correspondence:
\be \label{corres}
\partial_{\psi_\alpha} =  {e^{- i \nu_\alpha } \over 2 }  \left(\partial_{u_\alpha} -
{ i \over u_\alpha }  \partial_{\nu_\alpha} \right), \quad
\partial_{\bar \psi_\alpha} =  { e^{ i \nu_\alpha } \over 2 } \left( \partial_{u_\alpha} + { i \over u_\alpha }
\partial_{\nu_\alpha} \right), \quad \alpha=1,2.
\ee For example, we can write \be (\psi_\alpha
\partial_{\psi_\alpha} + {\bar \psi}_\alpha  \partial_{\bar
\psi_\alpha} ) = u_\alpha \partial_{u_\alpha}, \qquad  i
(\psi_\alpha \partial_{\psi_\alpha} -  {\bar \psi}_\alpha
\partial_{\bar \psi_\alpha} ) = \partial_{\nu_\alpha}, \quad
\alpha=1,2. \ee So from equations (\ref{chanvar}), (\ref{corres})
and after a slight change of basis, we get the following
generators: \beqa \nonumber X_1 &=& {1 \over 2 \wph} \sin 2\wph t
\partial_t + {x \over 2}  \cos 2\wph t \partial_x - {1 \over 4}  \cos
2\wph t
( u_1 \partial_{u_1} + u_2 \partial_{u_2} )   \\
\nonumber &-&  {M \wph x^2 \over 2} \sin 2\wph t  (
\partial_{\nu_1} +
\partial_{\nu_2}   )  +
{1\over 4} \sin 2\wph t   (\partial_{\nu_1} - \partial_{\nu_2} ),  \\
\nonumber X_2 &=& -{ 1 \over 2 \wph} \cos 2\wph t  \partial_t + {x
\over 2}  \sin 2\wph t \partial_x - {1 \over 4}  \sin 2\wph t
(   u_1 \partial_{u_1}  + u_2 \partial_{u_2})   \\
\nonumber &+&  {M \wph x^2 \over 2} \cos 2\wph t (
\partial_{\nu_1} +
\partial_{\nu_2}   )   -
{1\over 4} \cos 2\wph t   ( \partial_{\nu_1} - \partial_{\nu_2 }  ) , \\
\nonumber
X_3 &=& \partial_t + {\wph \over 2}  ( \partial_{\nu_1} - \partial_{\nu_2}   ) , \\
\nonumber
X_4 &=& \cos \wph t \partial_x -  M \wph x \sin \wph t    ( \partial_{\nu_1} + \partial_{\nu_2 }  ),  \\
\nonumber
X_5 &=& \sin \wph t \partial_x +   M \wph x \cos \wph t    ( \partial_{\nu_1}+ \partial_{\nu_2 }  ) , \\
\nonumber
X_6 &=&  ( \partial_{\nu_1} +  \partial_{\nu_2}   ), \\
\nonumber
X_7 &=& \cos (\wph t + {\nu_2} - {\nu_1} ) u_2 \partial_{u_1}  +  {u_2 \over u_1} \sin  (\wph t + \nu_2 - \nu_1 ) \partial_{\nu_1}, \\
\nonumber
X_8 &=& \cos (\wph t + \nu_2 - \nu_1 ) u_1 \partial_{u_2} -  {u_1 \over u_2} \sin  (\wph t + \nu_2 - \nu_1 ) \partial_{\nu_2} , \\
\nonumber
X_9 &=& u_2 \partial_{u_2}  - u_1 \partial_{u_1} , \\
\nonumber
X_{10} &=& \sin (\wph t + \nu_2 - \nu_1 ) u_2 \partial_{u_1}  -  {u_2 \over u_1} \cos (\wph t + \nu_2 - \nu_1 ) \partial_{\nu_1}, \\
\nonumber
X_{11} &=&  - \sin (\wph t + \nu_2 - \nu_1 ) u_1 \partial_{u_2} -  {u_1 \over u_2} \cos  (\wph t + \nu_2 - \nu_1 ) \partial_{\nu_2} , \\
\nonumber
X_{12} &= & \partial_{\nu_1} - \partial_{\nu_2}, \\
\nonumber
X_{13} &=&  u_1 \partial_{u_1}   + u_2 \partial_{u_2}.
\nonumber
\eeqa

Table \ref{tab:tablauno} shows the commutation relations between the
generators $X_j, \, j=1,2, \ldots,6.$ They form a Lie algebra
isomorphic to $sl(2, {\mathbb R} ) { \dotplus } h(2)=
\{X_1,X_2,X_3\} { \dotplus } \{X_4,X_5,X_6\} $. Table
\ref{tab:tablados} shows the commutation relations between the
generators $X_j, \, j=7, \ldots, 12, $ which form a Lie algebra
isomorphic to the complex extension of $su(2)$ denoted by
$su(2)^{\mathbb C}$. The generator $X_{13}$ is a central element in
this complete algebra. Since, the generators of table
\ref{tab:tablauno} commute with those of table \ref{tab:tablados},
we get the symmetry Lie algebra of the set (\ref{schrosusyuno}) as
$\{ sl(2, {\mathbb R} ) { \dotplus } h(2) \} \oplus su(2)^{\mathbb
C} \oplus \{X_{13}\}.$ The interpretation of these symmetries with
respect to other approaches require to compute the finite symmetry
transformations of the independent and dependent variables and also
a specific realization of the preceding generators. That's what we
propose to do in the following subsection.

\begin{table}[t]
\begin{center}
\begin{tabular}{|c|c|c|c|c|c|c|}\cline{2-7}
\multicolumn{1}{c|}{}& $X_1$ & $X_2$ &$X_3 $  & $X_4$ & $X_5$ &
$X_6$  \\  \cline{1-7} $X_1$ & $0 $ & $ {1\over 2\wph } X_3  $ & $
2\wph
X_2 $ & $ - {1\over 2}X_4  $ & $ {1\over 2} X_5  $ & $ 0 $  \\
\hline $X_2$ & $ - {1\over 2\wph} X_3   $ & $0$ &$ -2\wph X_1 $  &
$- {1\over 2}X_5 $ & $ - {1\over 2}X_4 $ & $0 $ \\  \hline $X_3 $
&
$- 2\wph X_2 $ & $2\wph X_1 $ &$0$ & $- \wph X_5 $ & $ \wph  X_4   $&$ 0 $ \\
\hline $X_4$ & $ {1\over 2}X_4  $ & ${1\over 2}X_5 $  & $ \wph X_5
$ & $0$ & $ M \wph X_6 $&$0$   \\ \hline $X_5$ & $- {1\over 2} X_5
$ &
${1\over 2} X_4 $ &$ - \wph X_4   $ & $ - M\wph X_6 $ &$ 0 $ & $0 $ \\
\hline $X_6$  & $0 $ & $ 0  $  & $ 0 $  & $ 0 $ & $0 $ & $0 $  \\
\hline
\end{tabular}
\end{center}
\caption{Commutation relations of a $sl(2, {\mathbb R}) \dotplus
 h(2)$ algebra.} \label{tab:tablauno}
\end{table}

\begin{table}
\begin{center}
\begin{tabular}{|c|c|c|c|c|c|c|}\cline{2-7}
\multicolumn{1}{c|}{}& $X_7$ & $X_{8}$ &$X_{9}$  & $X_{10}$ &
$X_{11}$ & $X_{12}$ \\ \cline{1-7}
$X_7$ & $0$ & $X_{9}$  & $-2 X_{7}$ & $0$ & $X_{12}$ & $-2 X_{10} $  \\
\hline
$X_{8}$ & $-X_{9}$ & $0$  & $2 X_{8}$ & $-X_{12}$ & $0$ & $2 X_{11}$  \\
\hline
$X_{9}$ & $2X_{7}$ & $-2X_{8}$  & $0$ & $2X_{10}$ & $-2X_{11}$ & $0 $ \\
\hline
$X_{10}$ & $0$ & $X_{12}$  & $-2X_{10}$ & $0$ & $-X_{9}$ & $2X_7$  \\
\hline
$X_{11}$ & $-X_{12}$ & $0$  & $2X_{11}$ & $X_{9}$ & $0$ & $-2X_{8} $ \\
\hline
$X_{12}$ & $2X_{10}$ & $-2X_{11}$  & $0$ & $-2X_{7}$ & $2X_{8}$ & $0$ \\
\hline
\end{tabular}
\end{center} \caption{Commutation relations of a complex extension of
$su(2).$} \label{tab:tablados}
\end{table}

\subsection{Integration of vector fields and realization of the generators}
Once we integrate the vector fields, we get the one parameter
groups of transformations which leave the equation
(\ref{schrosusyuno}) invariant. To the generator $X_1$, it
corresponds the following transformation (with the integration
parameter $\lambda_1$) on time and space coordinates
\be\label{t-x-uno} {\tilde t } = {1\over \wph} \arctan \left(
e^{\lambda_1} \tan \wph t \right) , \quad {\tilde x} =
e^{(\lambda_1 /2)} x { \left( { 1 + \tan^2 \wph t \over 1+ e^{2
\lambda_1 } \tan^2 \wph t  } \right)  }^{1/2} \ee and on the wave
function \beqa \label{psitilde-uno} {\tilde \Psi} (\tilde t ,
\tilde x ) &=& e^{\lambda_1 \over 4}  {\left( { 1 + \tan^2 \wph
\tilde t  \over  1+
e^{-2 \lambda_1 } \tan^2 \wph \tilde t  } \right)}^{1/4} \nonumber \\
& & \exp\left[  i   { M \wph  {\tilde x }^2 \over 2 \tan \wph
\tilde t }
\left( 1 - { 1 + \tan^2 \wph \tilde t  \over  1+ e^{- 2 \lambda_1 } \tan^2 \wph \tilde t } \right) \right]\nonumber \\
& & \pmatrix{e^{i\wph (\tilde t - t)}  & 0 \cr 0 &  e^{- i\wph
(\tilde t - t)} \cr}  \Psi  (t,x), \eeqa where $t$ and $x$ in the
expression of $\Psi (t,x)$ given before have to be evaluated using
the inverse of (\ref{t-x-uno}). To the generator $X_2,$ corresponds
\be \label{t-x-dos} {\tilde t } = {1 \over \wph} \arctan \left( e^{-
\lambda_2} \tan ({\pi \over 4} + \wph t ) \right) - {\pi \over
4\wph} , \qquad {\tilde x} =  e^{- {\lambda_2 \over 2}} x { \left( {
1 + \tan^2 ( {\pi \over 4} + \wph t ) \over 1+ e^{- 2 \lambda_2 }
\tan^2 ( {\pi \over 4} + \wph t ) } \right) }^{1/2} \ee and \beqa
{\tilde \Psi} (\tilde t , \tilde x ) &=& e^{\lambda_2 /4} {\left( {
1 + \tan^2 ({\pi \over 4} + \wph \tilde t )  \over  1+ e^{2
\lambda_2 }
\tan^2 ({\pi \over 4} + \wph \tilde t )  } \right)}^{1/4} \nonumber\\
& &\times \exp\left[  i   { M \wph  {\tilde x }^2 \over 2 \tan ({\pi
\over 4} +\wph \tilde t) } \left( 1 - { 1 + \tan^2 ({\pi \over 4} +
\wph \tilde t )  \over  1+ e^{2 \lambda_2 }
\tan^2 ( {\pi \over 4} + \wph \tilde t ) } \right) \right]\nonumber\\
& & \times \pmatrix{e^{i\wph (\tilde t - t)}  & 0 \cr 0 &  e^{-
i\wph (\tilde t - t)} \cr}  \Psi  (t,x), \eeqa where $t$ and $x$ in
the expression of $\Psi (t,x)$ in this equation have to be evaluated
using the inverse of (\ref{t-x-dos}). The generator $X_3$
corresponds to a time translation and the following transformation
of the wave function \be {\tilde \Psi} (\tilde t , \tilde x ) =
\pmatrix { e^{i\wph \lambda_3 /2} & 0 \cr 0 &
 e^{- i\wph \lambda_3 / 2} \cr} \Psi (\tilde t - \lambda_3 , \tilde x ).
\ee To the generator $X_4 $, we associate the transformation \be
\tilde t = t, \qquad  \tilde x =x + \lambda_4 \cos \wph t, \ee and
\be {\tilde \Psi} (\tilde t , \tilde x ) =  \exp\left[ - i M\wph
\left(  \lambda_4 \tilde x - {\lambda_4^2 \over 2} \cos \wph
\tilde t \right)  \sin \wph \tilde t \right] \Psi (\tilde t ,
\tilde x - \lambda_4 \cos \wph\tilde t  ). \ee The transformation
associated to the generator $X_5 $ is similar and gives \be \tilde
t = t, \qquad \tilde x =x + \lambda_5 \sin \wph t, \ee together
with \be {\tilde \Psi} (\tilde t , \tilde x ) =  \exp\left[  i
M\wph \left( \lambda_5 \tilde x - {\lambda_5^2 \over 2} \sin \wph
\tilde t \right) \cos \wph \tilde t \right] \Psi (\tilde t ,
\tilde x - \lambda_5 \sin \wph \tilde t  ) . \ee The generators
$X_j ,  \; j=6,7, \ldots, 13$ are not associated with space-time
transformations but with transformations of the wave function
which leave invariant the original set of equations. The
integration of these vector fields leads to the following
transformations: \beqa {\tilde \Psi} ( t , x ) &=& e^{i \lambda_6
}  \Psi  (t ,  x ), \qquad
{\tilde \Psi} (t ,  x ) = \pmatrix{1 & \lambda_7 e^{i \wph  t} \cr   0 &  1 \cr} \Psi  ( t ,   x ), \\
{\tilde \Psi} ( t ,  x ) &=& \pmatrix{1 & 0 \cr  \lambda_8 e^{- i
\wph t} & 1 \cr} \Psi  (  t ,   x ), \qquad
{\tilde \Psi} (t , x ) = \pmatrix{ e^{- \lambda_9 } & 0 \cr 0 & e^{ \lambda_9 } \cr} \Psi  ( t ,  x ), \\
{\tilde \Psi} (t ,  x ) &=& \pmatrix{1 & - i   \lambda_{10} e^{i
\wph t} \cr  0 & 1   \cr} \Psi  (  t ,   x ), \qquad
{\tilde \Psi} ( t ,  x ) = \pmatrix{1 & 0 \cr - i   \lambda_{11} e^{- i \wph  t}  & 1 \cr} \Psi  (  t ,   x ), \\
{\tilde \Psi} ( t ,  x ) &=& \pmatrix{ e^{i \lambda_{12} } & 0 \cr 0 & e^{- i \lambda_{12} }  \cr} \Psi  ( t ,  x ), \qquad
{\tilde \Psi} ( t , x ) = e^{\lambda_{13} }  \Psi  ( t ,   x ).
\eeqa
Now from these finite transformations we can find a matrix realization of the infinitesimal generators of the invariance Lie algebra.
It is easy to show that we get:
\beqa \label{decplus}
C_- (t) &=& 2\wph  (i  X_1- X_2 ) \nonumber \\
&=& e^{2i\wph t}  \left( (  \partial_t + i\wph x \partial_x  + i
M\wph^2 x^2 + i  {\wph \over 2} ) \sigma_0 -
  i { \wph \over 2} \sigma_3 \right) , \\
 C_+ (t) &=& -2\wph (i X_1+X_2 ) \nonumber \\
&=& e^{- 2i\wph t}  \left( (  \partial_t - i\wph x \partial_x
 + i  M\wph^2 x^2 - i  {\wph \over 2} ) \sigma_0  -  i { \wph \over 2} \sigma_3 \right), \\
H_0 &=& i X_3  = i \sigma_0 \partial_t +  {\wph \over 2}  \sigma_3,   \\
A_{x,-}  (t) &=& {1\over\sqrt{2M\wph}}  (X_4 + iX_5) =  { 1\over \sqrt{2M\wph}} e^{i\wph t} (M\wph x + \partial_x )  \sigma_0, \\
A_{x,+}  (t) &=& {- 1\over \sqrt{2M\wph}} (X_4 -  iX_5)  = { 1\over \sqrt{2M\wph}} e^{-i\wph t} (M\wph x - \partial_x )\sigma_0,    \\
I &= & i X_6  =  X_{13} = \sigma_0,  \\
\label{atplus}
\quad  T_+ ( t ) &=&   X_7  =-i X_{10}=  e^{i\wph t} \sigma_+ , \\ T_- ( t ) &=& X_8  =-i X_{11}= e^{-i\wph t} \sigma_- , \\
2 Y &=& X_9 = -i  X_{12} =  \sigma_3 ,\label{ygreque} \eeqa where $
\sigma_+ = \pmatrix{0&1 \cr 0&0\cr}= {1\over 2} (\sigma_1 + i
\sigma_2) $ and $ \sigma_- =\pmatrix{0&0 \cr 1&0\cr}= {1\over 2}
(\sigma_1 - i \sigma_2) , $ with $\sigma_1 , \sigma_2 , \sigma_3 $
the standard Pauli matrices. The generators $C_-,\ C_+,\ H_0 , \
A_{x,-}(t) ,\ A_{x,+}(t) $ and $I$ correspond exactly to the maximal
kinematical algebra $sl(2, {\mathbb R} ) {\dotplus } h(2)$. The
generators $T_+,\ T_-$ and $Y$ correspond to the algebra $su(2)$ and
are associated with the fermionic symmetries of the SUSY harmonic
oscillator.

Since we expect for the SUSY harmonic oscillator the presence of
bosonic (even) and fermionic (odd) symmetries, we can associate to
these generators a parity, i.e., those represented in terms of
diagonal matrices are called even and those represented by
anti-diagonal matrices are called odd. So they close now a Lie
superalgebra $$(sl(2, {\mathbb R} \dotplus sh(2/2)) \oplus \{Y\}.$$
The Lie superalgebra $sh(2/2 )$ is given by the set $\{ A_{x,-}(t) ,
\ A_{x,+}(t) , \, I ; \, T_+ (t), T_- (t)\} $ and the associated
non-zero super-commutation relations are \be [A_{x,-}(t)
,A_{x,+}(t)] = \{ T_- (t) , T_+ (t) \} = I. \ee

The SUSY generators are not obtained by these procedure. Since,
they play the role to exchange bosonic and fermionic fields, they
are know to be associated to a composition of even and odd
generators. Indeed, they may be written as the products: \beqa
\label{susyqq}
Q_+ &=& \sqrt{\wph} \ T_+(t)  A_{x,+}(t), \qquad Q_- =  \sqrt{\wph}\ T_- (t) A_{x,-}(t), \\
S_+ (t) &=& \sqrt{\wph} \ T_+ (t) A_{x,-}(t), \qquad S_-(t) =
\sqrt{\wph} \ T_-(t) A_{x,+}(t) \label{susyss} \eeqa and they close
together with the original ones, given in Eqs.
(\ref{decplus}-\ref{ygreque}), the superalgebra $osp(2/2){ \dotplus
} sh(2/2)$. So the prolongation method, using such matrix
realization has given bosonic and fermionic symmetries which close a
superalgebra. The SUSY generators present have been included by hand
by taking suitable products of some basic even $(A_{x,\pm} (t))$ and
odd ($T_\pm (t)$) generators. To explain why these products appear,
it is convenient to relate the prolongation method for searching
symmetries to the general concept of symmetry  of a quantum system.

We consider the general transformation  \be \label{tplusgen}
\tilde \Psi (t,x) = X \Psi(t,x) \ee on the wave function $\Psi
(t,x)$ of our quantum system (\ref{schopersusy}), where $X$ is a
operator  such  that \be \label{schoxmatrix} \left( i
\partial_t - H_{SUSY} \right) X \Psi(t,x) = 0, \ee
i.e., $X$ transforms solutions of our system into solutions. The
operator $X$ of (\ref{schoxmatrix}) is called a
symmetry operator of the model under study. In this more general context
it is clear that, if two operators $X$  and $Y$ satisfy (\ref{schoxmatrix}),
the product $XY $ also.
Moreover, if eq.(\ref{schopersusy}) satisfies the superposition principle
of solutions, the linear combination $\alpha X + \beta
Y,$ where $\alpha, \beta \in {\mathbb C}$ satisfies also
(\ref{schoxmatrix}). But  the complete set of operators obtained by this procedure
does not necessarily close a Lie algebra or superalgebra.

Let us take the
operator $X$ in (\ref{schoxmatrix}) to be on the differential
form\cite{kn:durand} \be \label{xmatrix} X = \sum_{\mu=0}^{4}
\left[ \sigma_\mu \left( \phi^0_\mu (t,x) + \phi^1_\mu (t,x)
\partial_t + \phi^2_\mu (t,x) \partial_x\right)\right],  \label{gentypex}\ee
where the $\phi^k_\mu (t,x), \ k=0,1,2; \mu=1,\ldots, 4, $ are
real functions of $t$ and $x.$ Comparing the coefficients of
several independent products of derivatives we get a system of
PDE's which can be solved to determine $\phi^0_\mu (t,x),
\phi^1_\mu (t,x) $ and $\phi^2_\mu (t,x),$ up to a finite
number of arbitrary integration constants. Solving this system,
inserting the results in (\ref{xmatrix}), identifying the
different operators according to each integration
constants, and finally,  taking suitable combinations of these
operators, we obtain the generators (\ref{decplus}-\ref{ygreque})
and also other ones. These last are the second order products of
the original ones, i.e., \beqa \label{deygriega}
Y C_+ (t), \quad Y C_-(t), \quad Y A_{x,+}(t), \quad Y A_{x,-} (t), \quad Y X_3 ,  \\
T_+ (t) C_+ (t), \quad T_+ (t) C_-(t), \quad T_+ (t) A_{x,+}(t), \quad  T_+  A_{x,-} (t), \quad T_+ (t) X_3, \\
\label{atmoins} T_- (t) C_+ (t), \quad T_- (t) C_-(t), \quad T_-
(t) A_{x,+} (t), \quad T_- A_{x,-} (t), \quad T_- (t) X_3. \eeqa
It is easy to show that the all set of symmetries does not close a
Lie algebra or a Lie superalgebra and
the only way to close the structure under both commutation and
anti-commutation relations is to select among all the preceding
products the four ones given in (\ref{susyqq}-\ref{susyss}).

Let us finally mention that on the space of solutions  of the
equation (\ref{schopersusy}), the superalgebra $osp(2/2) \dotplus
sh(2/2) $ may be expressed as  \beqa C_- (t) &=& i \wph e^{ 2i\wph
t} {a_x^2 \over 2} \sigma_0 , \quad C_+ (t) =  i \wph e^{-2i\wph t}
{{(a^\dagger_x )}^2 \over 2 }\sigma_0 , \quad
H_0 = \wph  \left( a^\dagger_x  a_x  + { 1\over 2} \right ),  \\
A_{x,-}(t) &=& e^{i\wph t} a_x \sigma_0, \qquad  A_{x,+}(t) =
e^{-i\wph t} a^\dagger_x \sigma_0 ,
 \qquad I = \sigma_0, \\
T_+ (t) &=& e^{i\wph t} \sigma_+, \qquad T_- (t) = e^{-i\wph t} \sigma_-, \qquad Y = {\sigma_3 \over 2},\\
Q_+ &=& \sqrt{\wph}\ a^\dagger_x \sigma_+ , \qquad Q_- = \sqrt{\wph} \ a_x \sigma_-, \\
 S_+ (t) &=& \sqrt{\wph} \ e^{2i\wph t} a_x \sigma_+ ,\qquad
S_-(t) =  \sqrt{\wph}\ e^{-2i\wph t} a^\dagger_x \sigma_- , \eeqa
where \be \label{annix} a_x = {1\over \sqrt{2M\wph}}( M\wph x +
\partial_x ) , \quad  a^\dagger_x = {1\over \sqrt{2M\wph}}( M\wph x -
\partial_x ) , \ee are the usual annihilation and creation
operators, respectively. They satisfy the commutation relation \be
[a_x, a^\dagger_x ] =1. \ee  Table \ref{tab:tablatres} shows the
commutation and anticommutation relations between the generators
of the orthosymplectic superalgebra and, as expected, the
generators $Q_\pm $ are the supercharges of the system. This means
that they satisfy \be
  \{ Q_+ , Q_- \} = H_0 - \wph Y = H_{SUSY} , \quad {(Q_\pm )}^2 =0
\ee and \be [H_{SUSY} , Q_\pm ] = 0 . \ee  Table
\ref{tab:tablacuatro} shows the structure relations between the
generators of $osp(2/2)$ and $sh(2/2)$.

\begin{table}[t]
\begin{center}
\begin{tabular}{|c|c|c|c|c||c|c|c|c|}\cline{2-9}
\multicolumn{1}{c|}{}& $ H_0 $ & $C_- (t)$ & $ C_+ (t) $
& $ Y $ & $Q_- $ & $Q_+ $ & $ S_- (t) $ & $ S_+ (t) $ \\
\cline{1-9} \hline $H_0 $ & $0$ & $ - 2\wph C_- $ & $ 2\wph  C_+ $
&$0$ & $-\wph  Q_- $ & $ \wph Q_+ $& $ \wph S_- $ & $ - \wph S_+ $ \\
\hline
$C_- (t) $ & $2\wph C_- (t) $ & $0$ & $- \wph H_0 $ & $0$ & $0 $& $ i \wph S_+ $ & $ i \wph Q_- $ & $0 $   \\
\hline $C_+ (t) $ & $- 2 \wph C_+ $ & $\wph H_0 $ &$0$ & $0$ & $
-i\wph  S_- $& $0 $ & $0$ & $-i\wph Q_+ $ \\ \hline $Y$ & $0$ &
$0$ & $0$ & $0$& $- Q_- $& $ Q_+ $ & $ - S_-  $ & $ S_+ $
\\ \hline \hline $Q_- $ & $\wph Q_- $ & $0$ &$ i\wph S_-  $ & $ Q_-
$ & $0 $ & $ H_0 - \wph Y $ & $0$ & $ -2 i C_- $  \\ \hline $Q_+ $
& $\wph Q_+ $& $ i\wph S_+ $ &$0$ & $- Q_+ $ & $H_0 - \wph Y$ &$0$
&
$-2i C_+ $ & $0$ \\
\hline $S_- (t) $ & $-\wph S_-  $ & $ - i \wph Q_- $ &$0$ & $2 S_-
$ &
$0$ &$-2i C_+ $ & $0$ & $H_0 + \wph Y$  \\
\hline $S_+ (t) $ & $\wph  S_+ $ & $0$ &$i \wph Q_+  $  & $-2 S_-  $ & $-2i C_- $ & $0$ & $H_0 + \wph Y $ & $0$ \\
\hline
 \end{tabular}
 \end{center}
\caption{Super-commutation relations of a $osp(2 /2 )$
superalgebra.} \label{tab:tablatres}
\end{table}

\begin{table}[t]
\begin{center}
\begin{tabular}{|c|c|c|c|c||c|c|c|c|}\cline{2-9}
 \multicolumn{1}{c|}{}&
$ H_0 $ & $C_- (t)$ & $ C_+ (t) $ & $ Y $ & $Q_- $ & $Q_+ $ & $
S_- (t) $ & $ S_+ (t) $ \\ \cline{1-9} $A_{x,-} (t) $ & $ \wph
A_{x,-} $ & $0$ &$ i\wph A_{x,+} $  & $0$ & $ 0 $& $\sqrt{\wph}
T_+ $ & $\sqrt{\wph} T_- $& $0$  \\ \hline $A_{x,+} (t) $ & $-
\wph A_{x,+} $ & $ -i\wph A_{x,-} $ & $0 $ & $0$& $- \sqrt{\wph}
T_- $ & $0 $& $0$& $- \sqrt{\wph} T_+  $  \\ \hline
$I $ & $0 $ & $ 0 $ &$0$ & $0$ & $0 $& $0$ & $0$ & $0 $  \\
\hline  \hline $T_- (t) $ & $0$ & $0 $ &$ 0$ & $T_-  $  & $0$ &
$\sqrt{\wph} A_{x,+} $ & $0$ & $ \sqrt{\wph} A_{x,-} $ \\ \hline
$T_+ (t) $ & $0$ & $0$ & $0$  & $- T_+ $ & $ \sqrt{\wph} A_{x,-} $
& $0$ & $ \sqrt{\wph} A_{x,+} $& $0 $
\\ \hline
 \end{tabular}
 \end{center}
\caption{Super-commutation relations between the generators of $
osp(2/2)$ and $sh(2/2).$} \label{tab:tablacuatro}
\end{table}

\section{A non-relativistic spin-${1\over 2}$ particle in a constant magnetic field}
\label{sec-dos} A problem which is related to the preceding one is
the search for symmetries and supersymmetries of the
Schr\"{o}dinger-Pauli equation describing the motion in the plane
of a non-relativistic spin-${1\over 2}$ particle of electric
charge $e$ in a constant magnetic field ${\vec B}= (0,0,B)$
orthogonal to the plane. We thus have the equation \be
\label{pauli-1} \left( i\partial_t - H_P \right) \Psi (t, x, y  )
= 0, \ee where $\Psi (t,  x,y  )$ is as given in
(\ref{wavefunction}) except that $\psi_1$ and $\psi_2$ depend on
$t,x$ and $y$.  The Hamiltonian is explicitly given by \be
\label{pauli-2} H_P = {{\left( { \vec \sigma} \cdot ( {\vec p} - e
\vec A )\right)}^2 \over 2M }= {{( {\vec p} - e \vec A )}^2 \over
2M} + i {\vec \sigma}
 \cdot ({ ( {\vec p} - e \vec A ) \times ( {\vec p} - e \vec A )}).
\ee
The vector $\vec \sigma$ is given by $\vec \sigma = (\sigma_1 , \sigma_2 , \sigma_3 ),$ where the $\sigma_i$'s are the usual Pauli matrices,
$\vec p = (p_x, p_y, 0 )$ is the linear momentum and
\be \label{potentialvector}
\vec A = ( - {1\over 2} B y
 \,  , \,  {1\over 2}B x \, , \, 0 )
\ee
 is the vector potential in the symmetric gauge . We can thus write the equation (\ref{pauli-1}) as
the following set of equations:
\be \label{2pauli}
\Biggl \{  i \partial_t  + {1 \over 2M}
\Biggl( {\partial^2 \over \partial x^2 } + {\partial^2 \over \partial y^2 } - i e B \left(x {\partial \over \partial y } - y {\partial \over \partial x }
\right)- {e^2 B^2 \over 4}(x^2 +y^2) + e B_\alpha )  \Biggr \} \psi_\alpha  (t,x,y) = 0,
\ee
with $\alpha=1,2$ and where we have set $ B_1 =B$, $B_2 = -B.$

To get the infinitesimal generators corresponding to the
symmetries of this set, we can again apply the prolongation method
where the wave function $\Psi (t,x,y)$ may be written now as \be
\label{psiu} \psi_1 (t,x,y) = u_1 (t,x,y) e^{i \nu_1 (t,x,y)},
\qquad \psi_2 (t,x,y) = u_2 (t,x,y) e^{i \nu_2 (t,x,y)}, \ee where
$ u_1, u_2, \nu_1 $ et $\nu_2 $ are real functions. The
corresponding vector field is \be \label{champ-vecteur} v=
\xi_1 \partial_t + \xi_2 \partial_x + \xi_3
\partial_y + \phi_1 \partial_{u_1} + \phi_2 \partial_{u_2} +
\varphi_1
\partial_{\nu_1} + \varphi_2 \partial_{\nu_2}, \ee
where $ \xi_j \ ( j=1,2,3),\ \phi_\alpha  $ and $\varphi_\alpha\ (
\alpha =1,2) $ are real functions of $t,x,y,u_1 ,u_2 , \nu_1,\nu_2
.$ As before, it is easier to make the calculation using the
complex form of the vector field \be \label{paulichampvec}
 v= \xi_1 \partial_t + \xi_2  \partial_x + \xi_3  \partial_y +
\Phi_1 \partial_{\psi_1}  + {\bar \Phi}_1 \partial_{\bar \psi_1} +
\Phi_2 \partial_{\psi_2}  + {\bar \Phi}_2 \partial_{\bar \psi_2},
\ee and after to go back to the real form. We finally get the
following generators, where we have introduced $ \wph = {eB\over
{2M}}$:
\begin{eqnarray*}
X_0 &=& \partial_t - \wph ( x \partial_y - y \partial_x ) + \wph (
\partial_{\nu_1 } - \partial_{\nu_2 }), \\ X_1 &=&  \cos2\wph t \partial_t - \wph ( x \sin 2\wph t - y \cos 2\wph t )
\partial_x - \wph ( x \cos 2\wph t + y \sin 2\wph t ) \partial_y  \\ &-& M \wph^2 (x^2 + y^2 ) \cos 2\wph t
(\partial_{\nu_1 } + \partial_{\nu_2 } ) + \wph ( \sin 2\wph t \;
(u_1
\partial_{u_1 } + u_2 \partial{u_2 } ) \\ &+& \cos 2\wph t \;
(\partial_{\nu_1 } - \partial_{\nu_2 } ) ), \\ X_2 &=& - \sin
2\wph t
\partial_t - \wph ( x \cos 2\wph t + y \sin 2\wph t ) \partial_x +  \wph ( x \sin 2\wph t - y \cos 2\wph t ) \partial_y  \\ &+& M \wph^2 (x^2 +
y^2 ) \sin 2\wph t (\partial_{\nu_1 } + \partial_{\nu_2 } ) + \wph ( \cos 2\wph t (u_1 \partial_{u_1 } + u_2 \partial{u_2 } ) \\ &-& \sin 2\wph t (\partial_{\nu_1 } - \partial_{\nu_2 } ) ), \\
X_3 &=& x \partial_y - y \partial_x, \\ X_4 &=& - {1 \over 2\wph}
(\cos 2\wph t \partial_x - \sin 2\wph t \partial_y ) + {M\over 2}
(x \sin
2\wph t + y \cos 2\wph t) ( \partial_{\nu_1 } + \partial_{\nu_2 } ), \\
X_5 &=& {1 \over 2\wph} (\sin 2\wph t \partial_x + \cos 2\wph t
\partial_y )
+ {M\over 2} (x \cos 2\wph t - y \sin 2\wph t ) ( \partial_{\nu_1 } + \partial_{\nu_2 } ), \\
X_6 &=& ( \partial_{\nu_1 } + \partial_{\nu_2 } ), \\
X_7 &=& \partial_x + M\wph  y ( \partial_{\nu_1 } + \partial_{\nu_2 } ), \\
X_8 &=& \partial_y - M\wph x ( \partial_{\nu_1 } + \partial_{\nu_2 } ), \\
X_9 &=& \cos( 2\wph t+ \nu_2 -\nu_1 ) u_2 \partial_{u_1 } + {u_2 \over u_1} \sin( 2\wph t+ \nu_2 - \nu_1 ) \partial_{\nu_1 }, \\
X_{10} &=& \cos( 2\wph t+ \nu_2 -\nu_1 )  u_1 \partial_{u_2 }  - {u_1 \over u_2 } \sin( 2\wph t+ \nu_2 - \nu_1 ) \partial_{\nu_2},\\
X_{11} &=& u_2 \partial_{u_2} - u_1 \partial_{u_1} \\ X_{12} &=&
\sin ( 2\wph t+ \nu_2 - \nu_1 )  u_2 \partial_{u_1 }  -
{u_2 \over u_1 } \cos ( 2\wph t+ \nu_2 - \nu_1 ) \partial_{\nu_1}, \\
X_{13} &=& - \sin ( 2\wph t+ \nu_2 - \nu_1 )  u_1 \partial_{u_2 }  - {u_1 \over u_2 } \cos ( 2\wph t+ \nu_2 - \nu_1 ) \partial_{\nu_1}, \\
X_{14} &=& \partial_{\nu_1 } - \partial_{\nu_2 }, \\ X_{15} &=& u_1
\partial_{u_1} + u_2 \partial_{u_2 },
\end{eqnarray*}
The commutation relations between the generators $X_\mu\
(\mu=0,1,2,\ldots, 8) $ are given in table \ref{tab:unavez} and give
an algebra isomporphic to $ \{sl(2, {\mathbb R} )\oplus so(2)\} {
\dotplus } h(4)$. It is easy to show that  the generators $X_\mu \
(\mu=9,10, \ldots, 14)$ form an algebra isomorphic to
$su(2)^{\mathbb C}$ and equivalent to the one given in table
\ref{tab:tablados} for the SUSY harmonic oscillator. These two sets
commute with each other and $X_{15} $ is a central element. So we
have an algebra isomorphic to $\{( sl(2,{\mathbb R}) \oplus so(2) ){
\dotplus } \ h(4)\}  \oplus su(2)^{\mathbb C} \oplus \{X_{15} \}.$

\begin{table}[t]
\begin{center}
\begin{tabular}{|c|c|c|c|c|c|c|c|c|c|}\cline{2-10}
 \multicolumn{1}{c|}{}& $X_0$ & $X_1$ &$X_2$  & $X_3$ & $X_4$ & $X_5$ & $X_6$ & $X_7$
& $X_8$ \\ \cline{1-10}
$X_0$ & $0 $ & $2 X_2$ & $-2\wph X_1$ & $0$ & $\wph X_5$& $-\wph X_4$ & $0$ & $\wph X_8 $ & $-\wph X_7 $  \\
\hline $X_1$ & $-2\wph X_2$ & $0$ &$-2\wph X_0$  & $0$ & $ {1\over
2}X_8 $&${1\over 2}X_7 $ & $0$ & $2\wph^2 X_5$ & $2\wph^2 X_4$
\\ \hline $X_2$ & $2\wph X_1$ & $2\wph X_0 $ &$0$ & $0$ & $-{1\over 2} X_7 $&${1\over 2} X_8 $ & $0 $ & $- 2\wph^2 X_4$ & $2\wph^2 X_5 $\\
\hline $X_3$ & $0$ & $0$ & $0$  & $0$ & $X_5$&$-X_4$ & $ 0$ & $-X_8$ & $X_7$  \\ \hline
$X_4$ & $-\wph X_5$ & $- {1\over 2} X_8 $ &${1\over 2}X_7 $& $- X_5$ & $0 $ & $-  {M\over 2 \wph}  X_6 $ & $0$ & $0$ & $0  $ \\
\hline $X_5$ & $\wph X_4$
& $-{1\over 2} X_7 $ &$-{1\over 2} X_8 $ & $X_4$ & ${M \over 2 \wph}  X_6 $ &$0$ & $0$ & $0$ & $0 $\\
\hline $X_6$ & $0$ & $0$ &$0$ & $0$ & $0$&$0$ & $0$ & $0$ & $0$ \\
\hline $X_7$ & $-\wph X_8$ &  $- 2\wph^2 X_5$ &$2 \wph^2 X_4$  &
$X_8 $ & $0$ & $0$ & $0$ & $0$ &$-2 M \wph X_6 $ \\ \hline $X_8$ &
$ \wph X_7$ & $- 2\wph^2 X_4$ &$- 2 \wph^2 X_5$  & $-X_7$ & $0$ &
$0$ & $0$ & $2M\wph X_6$ & $0 $ \\ \hline
\end{tabular}
\end{center} \caption{Commutation relations for a $ \{sl(2, {\mathbb
R} )\oplus so(2)\} { \dotplus } h(4)$ algebra. } \label{tab:unavez}
\end{table}

Once again, we can get a specific matrix realization proceeding as
in the case of the SUSY harmonic oscillator. It is easy to show
that we have the following form for the generators of symmetries
of the equation (\ref{pauli-1}):

\beqa
H_0 & =& i X_0 =  \left( i \partial_t + \wph (xp_y - yp_x) \right) \sigma_0 + \wph \sigma_3, \\
\nonumber C_- (t) &=& {(X_1 - i X_2 ) \over 2} ={ e^{2i\wph t}\over
2} \{ (
\partial_t - i \wph ( x
p_y - y p_x) \\
&-& \wph (x p_x + y p_y)  + i \wph  + i M \wph^2 (x^2 + y^2 ))
\sigma_0 - i \wph \sigma_3 \},\\ \nonumber C_+ (t) &=& {(X_1 + i X_2
) \over 2 } = {e^{-2i\wph t} \over 2} \{ (
\partial_t - i \wph ( x p_y - y p_x)
\\ &+& \wph (x p_x + y p_y)  - i \wph  + i M \wph^2 (x^2 + y^2 )) \sigma_0  - i \wph
\sigma_3 \}, \\
L &=& - i X_3 = xp_y - y p_x, \\
\label{genpaulia}
{\cal A} (t) &=&  \sqrt{\wph \over M}(i X_4 + X_5) = {e^{2i\wph t} \over 2 \sqrt{\wph  M}} [( p_x + i p_y )- i M\wph (x + i y )] \sigma_0, \\
\label{genpauliad}
{\cal A}^\dagger (t) &=&  \sqrt{\wph \over M}(i X_4 - X_5) = {e^{- 2i\wph t}\over 2 \sqrt{\wph  M}}  [(p_x - i p_y )+ i  M\wph ( x - i y )] \sigma_0, \\
I &=& X_6 = X_{15}= \sigma_0, \\
\label{genpauliaa}
A_-  &=& {- 1\over 2 \sqrt{M\wph}} (X_8 + iX_7 ) = {1\over 2 \sqrt{M\wph}}[(p_x - i p_y )- i  M\wph ( x - i y  )] \sigma_0, \\
\label{genpauliaad}
A_+  &=&{1\over 2 \sqrt{M\wph}}(X_8 - iX_7)  = {1\over 2 \sqrt{M\wph}}[(p_x + i p_y  )+ i  M\wph (x  + i y )] \sigma_0, \\
T_+ (t ) &= &X_9 =-iX_{12}= e^{2i\wph t} \sigma_+ , \qquad T_- (t) = X_{10} =-iX_{13}=e^{-2i\wph t} \sigma_- , \\
 Y  &=&  X_{11} =-i X_{14}= \sigma_3.
\eeqa

We see that $H_0$ is essentially the Hamiltonian of the harmonic
oscillator in two dimensions, $C_\pm (t)$ correspond to the
so-called conformal transformations and $L$ is the angular momentum.
The two sets $\{ {\cal A} (t)$, \ ${\cal A}^\dagger (t)\}$ and
$\{A_- $,\ $A_+ \}$ may be associated to pairs of annihilation and
creation operators and $I$ represents the identity generator.
Indeed, they can be written as \be {\cal A} (t) = {1\over \sqrt{2}}
e^{2i\wph t} (a_y-i a_x) \sigma_0, \qquad {\cal A}^\dagger (t) =
{1\over \sqrt{2}} e^{-2i\wph t} (a_y^\dagger+i a_x^\dagger) \sigma_0
, \ee \be A_- =-{1\over \sqrt{2}} (a_y+i a_x) \sigma_0 , \qquad A_+
=-{1\over \sqrt{2}} (a_y^\dagger-i a_x^\dagger) \sigma_0 , \ee where
$a_x,\ a_x^\dagger,\ a_y$ and $\ a_y^\dagger$ are defined as in
(\ref{annix}). The set $\{ T_+ (t), \, T_- (t),Y\}$ corresponds to
the Lie algebra $su(2)$. These generators form the maximal
kinematical algebra of the Pauli equation (\ref{pauli-1}) which is
$\{ (sl(2,{\mathbb R} )  \oplus so(2))  { \dotplus } h(4) \}\oplus
su(2).$

Now the products of the generators which will lead to the
invariance superalgebras are obtained from the sets $\{ {\cal A}
(t) $,\ ${\cal A}^\dagger (t)\}$, $\{ A_- $,\ $A_+ \}$ and $\{T_+
(t), \, T_- (t)\}$. There are eight possible products. If we first
take \be  \label{susyQ} Q_- = \sqrt{2 \wph}\ {\cal A} (t) \, T_-
(t), \qquad Q_+  = \sqrt{2 \wph}\ {\cal A}^\dagger (t) \, T_+ (t),
\ee we see that they are independent on time as it was the case
with the SUSY harmonic oscillator. Moreover they satisfy \be Q_-^2
= Q_+^2 = 0 \ee and \be \{Q_- , Q_+ \} = H_P = H_0 - \wph L - \wph
Y, \qquad [H_P ,Q_\mp ] =0. \ee This means that $Q_\mp$ are the
supercharges for the Pauli Hamiltonian $H_P$ and  a class of
supersymmetries of our system. Another set of supersymmetries is
found to be \be S_- (t) = \sqrt{2\wph}\ A_+ \ T_- (t), \qquad S_+
(t) = \sqrt{2\wph} \ A_- \ T_+ (t). \ee Let us insist on the fact
that the two sets of annihilation and creation operators $\{ {\cal
A} (t)$, \ ${\cal A}^\dagger (t)\}$ and $\{A_- $,\ $A_+ \}$ thus
appear in these supersymmetry generators. The operators $S_\pm
(t)$, which are now time dependent, satisfy \be {(S_- (t))}^2 =
{(S_+(t))}^2 = 0 \ee and \be \{S_- (t), S_+ (t)\} = H_0 + \wph L +
\wph Y, \qquad [i\partial_t - H_P , S_\pm (t) ] = 0. \ee
 The other structure relations are computed and the non-zero ones are
\beqa
\displaystyle [ H_0 , Q_\pm ] = \pm  \wph Q_\pm &,& \qquad  [ H_0 , S_\pm ] = \mp  \wph S_\pm (t),  \\
\displaystyle [ C_+ (t ) , Q_- ] =  i \wph S_- (t) &,& \qquad [ C_- (t) , Q_+ ] = - i\wph S_+ (t), \\
\displaystyle [ Y , Q_\pm  ] =\pm 2 Q_\pm  &,& \quad [ Y , S_\pm (t) ] = \pm 2 S_\pm (t),\\
\displaystyle \{Q_- , S_+ (t) \} =  2 i C_- (t) &,& \qquad \{Q_+ ,
S_- \} = 2 i C_+ (t). \eeqa

This means that the generators $\{H_0, \ C_\pm (t), \ Y, \ Q_\pm, \
S_\pm (t)\}$ close the orthosymplectic superalgebra $osp(2/2).$
Together with the other generators of the maximal kinematical
algebra, we get the so-called maximal kinematical superalgebra of
the Pauli equation defined as $\{ osp(2/2) \oplus so (2)\} {
\dotplus} sh(2/4 ),$  where $sh(2/4)$ is the Heisenberg-Weyl
superalgebra generated by the fermionic generators $T_\pm (t)$, the
bosonic generators ${\cal A} (t), {\cal A}^\dagger (t), A_- , A_+ $
and the identity $I$ \cite{kn:BDH}. Let us finally mention that the
remaining products of $\{ {\cal A} (t)$, \ ${\cal A}^\dagger (t)\}$
and $\{A_- $,\ $A_+ \}$ with $\{T_+(t), \, T_- (t)\}$ give rise to
the following generators
\beqa U_+ (t) = \sqrt{2\wph}A_+  \, T_+ (t), \qquad U_- (t) =  \sqrt{2\wph} A_- \, T_- (t), \\
V_+ (t) =\sqrt{2\wph}  {\cal A} (t) \, T_+ (t), \qquad V_- (t) =
\sqrt{2\wph} {\cal A}^\dagger (t) \,  T_- (t). \eeqa They have been
introduced in \cite{kn:BDH} and are contained in the maximal
dynamical superalgebra of the system under consideration which is
$osp(2/4) { {\ensuremath{\rlap{\raisebox{.15ex}{$ \mskip 6.0
mu\scriptstyle+ $ }}\supset}\,\,} } sh(2/4 )$. Indeed to close the
structure with these additional supersymmetries, it is necessary to
include new even generators of the dynamical algebra of our model.

\section{The Jaynes-Cummings model}

Now we consider the system described by a particle of electric
charge $e,$  spin $1\over 2$ and mass $M$ moving in the plane in
the presence of constant electric $\vec E $ and magnetic $\vec B $
fields which are both perpendicular to the plane. It has been shown to be related to the Jaynes-Cummings model \cite{kn:GrLa}.
Indeed the Hamiltonian characterising such a system is given by \be H = {{\left( \vec p -
e \vec A \right)}^2  \over 2M} - { e \over 2M} \vec B  \cdot \vec
\sigma + {e  \over 4M^2 } {\vec E} \cdot
 \left(\vec  \sigma \times  ( \vec p - e \vec A ) \right)  + e \vec E \cdot \vec x,
\ee
where $\vec x$ is the position vector of the particle and $\vec A$ is the potential vector given in (\ref{potentialvector}). We are again
interested in the motion in the $xy$-plane for which the contribution of the  preceding Hamiltonian is
\beqa \label{hjc-complet}
H_{JC} &=& {1 \over 2 M} (  p^2_x + p^2_y + {e^2 B^2 \over 4} (x^2 + y^2) + eB (y p_x - x p_y ) ) \sigma_0  -
{eB\over 2M} \sigma_3 \\  &+&  {ie E \over 4 M^2 } \left(  (p_x - i p_y  ) + i {e B \over  2} ( x -i y) \right) \sigma_+   -
  {ie E \over  4 M^2 } \left(  (p_x + i p_y  ) -  i {e B \over  2} ( x +i y) \right) \sigma_- ,  \nonumber
\eeqa
or again,
\be
H_{JC} = H_P +  {ie E \over 4 M^2 } \left(  (p_x - i p_y  ) + i {e B \over  2} ( x -i y) \right) \sigma_+  -
  {ie E \over  4 M^2 } \left(  (p_x + i p_y  ) - i {e B \over  2} ( x +i y) \right) \sigma_- ,
\ee where $H_P$ is the Pauli Hamiltonian (\ref{pauli-2}). Let us
assume without loss of generality that $e > 0$ and introduce the
operators \beqa \label{new-cre} {\cal A} &=&{\cal A}(0)= {1\over
\sqrt{2eB} } \biggl(  (p_x + i p_y) - i{ e B\over 2}  (x+ i y)
\biggr) \sigma_0 ,\\ \label{new-ann} {\cal A^\dagger}&=&{\cal
A^\dagger}(0) ={1\over \sqrt{2eB} }\biggl( (p_x - i p_y) + i{ e
B\over 2} (x-  i y) \biggr)\sigma_0 , \eeqa which satisfy the
commutation relation \be [{\cal A},{\cal A^\dagger}] = I. \ee
These operators are nothing else than the generators given by
(\ref{genpaulia}) and (\ref{genpauliad}) at $t=0$ when we take
again $\wph={eB\over {2M}}$. Let us recall that they correspond to
symmetries of the Pauli system. Here we will see that they are not
symmetries of the JC model. The Hamiltonian $H_{JC}$ can thus be
written on the form \be \label{hajainescu} H_{JC} = {\tilde \wph}
\left( {\cal A^\dagger} {\cal A}  +  {1 \over 2} \right) \sigma_0
- {{\tilde \wph} \over 2} \sigma_3 + \kappa {\cal
A^\dagger}\sigma_+ +  {\bar \kappa } {\cal A} \sigma_-, \ee where
${\tilde \wph} = 2 \wph= {eB \over M} $ and $\kappa = {i eE
\sqrt{2eB} \over {4 M^2}} . $ This means that the Hamiltonian
(\ref{hajainescu}) is a realization of the JC Hamiltonian
\cite{kn:JaCu, kn:GrLa, kn:BHN} in the special case where the
detuning between the frequency of the cavity mode and the atom
transition frequency is zero. We also see a close connection with
the SUSY harmonic oscillator Hamiltonian described in terms of new
annihilation and creation operators ${\cal A}$ and ${\cal
A^\dagger}$ as given in (\ref{new-cre}-\ref{new-ann}).

\subsection{Lie algebra of symmetries}
We are interested to determine the symmetries of the corresponding evolution equation
\be \label{jctypeschro}
(i \partial_t - H_{JC} ) \Psi (t,x,y) =0,
\ee
where $\Psi (t,x,y)$ is again a two component wave function as in the Pauli equation
considered in the preceding section. It can be written explicitly
\beqa
  \label{systeme-jc}
  \nonumber
  i \psi_{1,t} &+& {1 \over 2M} \left[  \psi_{1,xx} + \psi_{1,yy} -   ieB (x\ \psi_{1,y} - y\ \psi_{1,x})  -
  { e^2 B^2 \over 4} (x^2 + y^2)\ \psi_1 +  e B_1 \psi_1 \right] \\ &+&  { eE_1 \over 4M^2} \left( \psi_{2,x} + i {e B \over 2} y\ \psi_2 \right) +
 { e E \over 4M^2 } \left( i \psi_{2,y} + {e B \over 2} x \ \psi_2  \right) = 0,  \label{systeme-jcuno}   \\
 \nonumber i \psi_{2,t} &+& {1 \over 2M} \left[  \psi_{2,xx} + \psi_{2,yy} -   ieB (x \ \psi_{2,y} - y \ \psi_{2,x})  - { e^2 B^2
 \over 4} (x^2 + y^2) \ \psi_2 +  e B_2  \psi_2 \right]  \\ & + &  { e E_2 \over 4M^2} \left( \psi_{1,x} + i {e B \over 2} y\ \psi_1
  \right) + { e E \over 4M^2 } \left(  i \psi_{1,y}+ {e B \over 2} x\ \psi_1  \right) = 0,  \label{systeme-jcdos}
\eeqa
where we have set $B_1=B, \, B_2 = - B, \, E_1 = - E$ and $E_2 = E.$

As in the preceding sections, to get the symmetries of the system
(\ref{systeme-jcuno}-\ref{systeme-jcdos}), we apply the
prolongation method to it and the conjugated system. Once again
the vector field has the form (\ref{paulichampvec}) with
\be \xi_1 = \delta_1, \qquad \xi_2 (t,y) = - \delta_2 y +
\delta_3, \qquad \xi_3(t,x) = \delta_2 x + \delta_4, \ee \be
 \Phi_1(t,x,y) = A_0 (t, x,y) + A_1 (x,y) \psi_1, \ \Phi_2 (t,x,y) = C_0 (t,x,y) +  C_2 (x,y) \psi_2,
 \ee
where
 \beqa
 A_1 (x,y) &=& - i { e B \over 2} (\delta_4  x - \delta_3 y )  - i { \delta_2  \over 2} + ( \delta_6 + i \delta_5 ), \\
  C_2 (x,y) &=& - i { e B \over 2} (\delta_4  x - \delta_3 y )  + i { \delta_2  \over 2} + ( \delta_6 + i \delta_5 ).
 \eeqa
The $\delta_j \ ( j=1,\ldots 6)$ are arbitrary real constants and
$A_0 (t, x,y)$ and $C_0 (t, x,y)$ are arbitrary functions that
satisfy the system (\ref{systeme-jcuno}-\ref{systeme-jcdos}) for
$\psi_1=A_0$ and $\psi_2=C_0$ . The finite  dimensional Lie
algebra of symmetries is thus formed by the following
infinitesimal generators: \beqa \nonumber
 X_1 &=& \partial_t, \\
\nonumber
 X_2 &=& (x \partial_y - y \partial_x ) - {i \over 2} (\psi_1 \partial_{\psi_1} - {\bar \psi}_1 \partial_{\bar \psi_1} ) +
 {i \over 2} (\psi_2 \partial_{\psi_2} - {\bar \psi}_2 \partial_{\bar \psi_2} ) , \\
\nonumber
 X_3 &=& \partial_x + i {eB \over 2} y \ (\psi_1 \partial_{\psi_1} - {\bar \psi}_1 \partial_{\bar \psi_1} + \psi_2 \partial_{\psi_2} -
{\bar \psi}_2 \partial_{\bar \psi_2} ) , \\
\nonumber
 X_4 &=& \partial_y -  i {eB \over 2} x \ (\psi_1 \partial_{\psi_1} - {\bar \psi}_1 \partial_{\bar \psi_1} + \psi_2 \partial_{\psi_2} -
 {\bar \psi}_2 \partial_{\bar \psi_2} ) , \\
\nonumber
 X_5 &=& i  (\psi_1 \partial_{\psi_1} - {\bar \psi}_1 \partial_{\bar \psi_1} ) + i ( \psi_2 \partial_{\psi_2} -
{\bar \psi}_2 \partial_{\bar \psi_2} ),  \\
\nonumber
 X_6 &=&  (\psi_1 \partial_{\psi_1} + {\bar \psi}_1 \partial_{\bar \psi_1} ) +
 ( \psi_2 \partial_{\psi_2} +  {\bar \psi}_2 \partial_{\bar \psi_2} ).
\nonumber
 \eeqa

 Using the real components of the wave function $\Psi (t,x,y)$ given by (\ref{psiu}), the infinitesimal generators take the form
 \beqa \label{algebrajc}
 X_1 &=& \partial_t,
\label{algebra1} \\
 X_2  &=& (x \partial_y - y \partial_x ) - {1\over 2} ( \partial_{\nu_1} -  \partial_{\nu_2} ),
\label{algebra2} \\
 X_3 &=&  \partial_x  + {e B\over 2} y  ( \partial_{\nu_1} + \partial_{\nu_2} ),
\label{algebra3} \\
 X_4 &=&  \partial_y  - {e B\over 2} x  ( \partial_{\nu_1} + \partial_{\nu_2} ),
\label{algebra4} \\
 X_5 &=&  ( \partial_{\nu_1} + \partial_{\nu_2} ),
\label{algebra5} \\
 X_6 &=&  u_1 \partial_{u_1} + u_2 \partial_{u_2} .
\label{algebra6}
 \eeqa

It is easy to see that $X_1$ and $X_6$ are both central elements.
The generator $X_2$ corresponds to a $so(2)$ algebra and the set
$\{X_3, X_4, X_5\}$ generates $h(2)$. These last generators are thus
associated to a Lie algebra isomorphic to $  so(2) { \dotplus } h(2)
$ and satisfy the following non-zero commutation relations \be
[X_2,X_3]=- X_4,\qquad [X_2,X_4]= X_3,\ee \be [X_3,X_4]= -eB X_5.
\ee

Integration of the vector fields gives rise to finite transformations of independent and dependent variables which leave the
equation (\ref{jctypeschro}) invariant. Explicitly, to $X_1$ corresponds the invariance under time translation such that:
\be
 {\tilde \Psi} ( {\tilde t}, {\tilde x}, {\tilde y}) = \Psi (\tilde t - \lambda_1, {\tilde x}, {\tilde y} ).
\ee

 The vector field $X_2$ corresponds to the invariance under rotations in the $xy$-plane. The transformation is
 \be
 ({\tilde t }, {\tilde x}, {\tilde y} )  =(t, x \cos \lambda_2 - y \sin \lambda_2, x \sin \lambda_2 +  y \cos \lambda_2)
 \ee
and
 \be
 {\tilde \Psi} ( {\tilde t}, {\tilde x}, {\tilde y}) =\pmatrix{e^{-{i\over 2} \lambda_2 } & 0 \cr 0 &  e^{{i\over 2} \lambda_2 } \cr}\
 \Psi ( {\tilde t},{\tilde x} \cos \lambda_2 + {\tilde y} \sin \lambda_2,  - {\tilde x} \sin \lambda_2 + {\tilde y} \cos \lambda_2 ).
 \ee

The vector fields $X_3 $ and $X_4$ correspond to the invariance under space translations. We have
\be
{\tilde \Psi} ( {\tilde t}, {\tilde x}, {\tilde y}) =
e^{i{e B\over 2} \lambda_3 {\tilde y}} \ \Psi (\tilde t , {\tilde x} - \lambda_3 , {\tilde y} )
\ee
 and
 \be
{\tilde \Psi} ( {\tilde t}, {\tilde x}, {\tilde y}) =
e^{- i{e B\over 2} \lambda_4 {\tilde x}}\  \Psi (\tilde t , {\tilde x} , {\tilde y} - \lambda_4 ),
 \ee
respectively. The vector fields $X_5$ and $X_6 $ are not related
to space-time transformations but to the following phase and
 scale transformations of the wave function respectively:
 \be
 {\tilde \Psi} ( t , x , y ) = e^{i \lambda_5} \Psi (t , x , y)
 \ee
and
 \be
 {\tilde \Psi} ( t , x , y ) = e^{ \lambda_6}  \Psi (t , x , y) .
 \ee
From these finite transformations, we easily get a matrix realization of the Lie algebra of symmetries of the equation (\ref{jctypeschro}).
 \beqa
 X_1 &=& \sigma_0 \partial_t, \qquad X_2 = (x \partial_y - y \partial_x ) \sigma_0 + { i \over 2} \sigma_3,  \\
 X_3 &=& \left( \partial_x - i {eB \over 2} y \right) \sigma_0, \qquad X_4 = \left( \partial_y +  i {eB \over 2} x \right)  \sigma_0, \\
 X_5 &=& -i X_6= - i I= - i \sigma_0.
\eeqa
Now $X_1$, while acting on the space of solution of (\ref{jctypeschro}), is the Hamiltonian $H_{JC}$ as expected and $J=-i X_2$ is the total
angular momentum. Complex linear combinations of $X_3$ and $X_4$ give
 \be \label{anni-jc}
A_- = { 1 \over
 \sqrt{2 e B} } \left[ (p_x - i p_y ) - i { e B \over 2} ( x - i y
 ) \right] \sigma_0 \ee
and
 \be \label{crea-jc}
  A_+ = { 1 \over  \sqrt{2 e B} } \left[  (p_x + i p_y ) + i { e B \over 2} ( x + i y  )  \right]  \sigma_0
 \ee
which satisfy \be [A_- , A_+ ] = I. \ee They are exactly the
symmetries of the Pauli Hamiltonian as given in (\ref{genpauliaa})
and (\ref{genpauliaad}) and close together with the identity $I$ the
algebra $h(2)$. The operators $\{H_{JC}, J, A_\mp , I\}$ are thus
associated to the maximal kinematical invariance algebra of the JC
model and is isomorphic to $ (so(2) { \dotplus } h(2) ) \oplus u(1).
$ They are all time independent and thus commute with $H_{JC}.$
Moreover they are all diagonal matrices and correspond necessarily
to even generators.

It is neither possible from the prolongation method to produce a
Lie superalgebra of symmetries nor a set of supersymmetries as it
was the case for the preceding models. We know in fact
\cite{kn:AnLe, kn:BuRa} that the standard JC model does not admit
a N=2 supersymmetry. It is due to the presence in the Hamiltonian
(\ref{hajainescu}) of the additional term $(\kappa {\cal
A^\dagger}\sigma_+ +  {\bar \kappa } {\cal A} \sigma_-)$ which can
be written as \beqa {\cal Q}&=&{1\over \sqrt{2 \wph}} ( \kappa
{Q_+}+{\bar \kappa} {Q_-})=\kappa {\cal A^\dagger}\sigma_+ + {\bar
\kappa } {\cal A} {\sigma_-}, \eeqa where  ${Q_+}$ and ${Q_-}$ are
given in (\ref {susyQ}). Due to the value of $\kappa$, such a term
is essentially a multiple of the combination $Q_+-Q_-$. It is an
additional symmetry of $H_{JC}$ which can not be obtained from the
prolongation method. It commutes with $A_\pm,\ \sigma_0$ and $J$
so that the set $\{H_{JC},A_\pm, I, J,\ Q_+-Q_-\}$ close a Lie
algebra but not a Lie superalgebra. Indeed, $(Q_+-Q_-)^2$ is not a
linear combination of the preceding generators.

\section{A generalized Jaynes-Cummings model}
As already mentioned in the approach by Andreev and Lerner
\cite{kn:AnLe}, to be able to get a supersymmetry for the JC
model, it is necessary to consider a $4$ by $4 $ matrix version of
the JC Hamiltonian. Let us show here how the prolongation method
may be adapted to such a model and will lead to the presence of
supersymmetry generators as for the SUSY harmonic oscillator and
the Pauli Hamiltonians. We first construct a generalized JC model
for which the prolongation method will produce symmetries
associated with a Lie superalgebra. Next we examine the
possibility to get supersymmetries for a symmetrized JC model and
compare the results with the ones associated with the so-called
standard SUSY JC model \cite{kn:OrSa}.

Let us start with a version of the JC Hamiltonian of the following
type: \be \label{hatotal} H_T=\pmatrix{H_{JC} - {eB \over
2M}\alpha \sigma_0  &  0 \cr 0 &
 H_{JC} -  {eB\over 2M} \beta \sigma_0 \cr},
\ee where $H_{JC} $ is given in (\ref{hjc-complet}) and $\alpha,
\beta $ are real parameters. The Schr\"odinger type equation is
\be \label{Schrot} (i \partial_t - H_T ) \Psi(t,x,y ) =0, \ee
where $\Psi(t,x,y)$ is a four component wave function whose entries are
complex valued functions equal to $\psi_\rho (t,x,y)\ ( \rho =1,
\ldots , 4)$. Since  (\ref{hatotal}) is diagonal we get a system
of four equations which are firstly given by
(\ref{systeme-jc}-\ref{systeme-jcdos}), where now we have set
$B_1=(\alpha + 1) B $ and $B_2 =( \alpha - 1) B$ while $E_1,\ E_2$
are still given by $E_1=-E_2=-E$. The second set of equations is
similar and we have \beqa  \nonumber i \psi_{3,t} &+& {1 \over 2M}
\left[ \psi_{3,xx} + \psi_{3,yy} - ieB (x\ \psi_{3,y} - y\
\psi_{3,x})  - { e^2 B^2 \over 4} (x^2 + y^2) \psi_3 +  e B_3
\psi_3 \right] \\ &+& { eE_3 \over 4M^2} \left( \psi_{4,x} + i {e
B \over 2} y\ \psi_4 \right) + { e {\tilde E}_3 \over 4M^2 }
\left( i \psi_{4,y}
+ {e B\over 2} x \ \psi_4 \right) = 0, \label{jctres}  \\
\nonumber
i\psi_{4,t} &+& {1 \over 2M} \left[ \psi_{4,xx} +\psi_{4,yy} - ieB (x \ \psi_{4,y} - y \ \psi_{4,x})  -
{ e^2 B^2 \over 4} (x^2 + y^2) \psi_4 +  e B_4 \psi_4 \right]  \\
& + &  { e E_4 \over 4M^2} \left( \psi_{3,x} + i {e B \over 2} y\
 \psi_3 \right) + { e {\tilde E}_4 \over 4M^2 } \left( i \psi_{3,y}+ {e B \over 2} x \ \psi_3 \right) = 0,
\label{jccuatro} \eeqa where we have set $B_3=(\beta + 1) B , \,
B_4 = (\beta -1) B , \, E_3 = - E_4 =- E $ and $ {\tilde E}_3 =
{\tilde E}_4 = E. $ The prolongation method applied to this system
and the associated complex conjugated equations leads to a vector
field of the form \be \label{jcchampvect} v = \xi_1 \partial_t +
\xi_2
\partial_x + \xi_3 \partial_y +
 \sum_{\rho=1}^{4} \Phi_\rho \partial_{\psi_\rho}  +
 \sum_{\rho=1}^{4} {\bar \Phi}_\rho \partial_{{\bar \psi}_\rho }, \ee
where $\xi_j \ ( j=1,2,3) $ and  $\Phi_\rho \ (\rho=1,\ldots, 4) $
are functions dependent on $t , x, y, \psi_\rho $ and ${\bar \psi}
_\rho . $ We get the solutions \be \label{xi-jc} \xi_1 = \delta_1,
 \qquad \xi_2 (t,y) = - \delta_2 y + \delta_3, \qquad \xi_3(t,x) =
 \delta_2 x + \delta_4,
\ee
\beqa \label{granphiuno}
\Phi_1(t,x,y) &=& A_0 (t, x,y) + A_1 (x,y) \psi_1 + f (t) \psi_3,\\
\label{granphidos}
\Phi_2(t,x,y) &=& C_0 (t,x,y) + C_2 (x,y) \psi_2 +  f (t) \psi_4,\\
\label{granphitres}
\Phi_3 (t,x,y)&=& D_0 (t,x,y) + g (t) \psi_1  + D_3 (x,y)\psi_3, \\
\label{granphicuatro}
\Phi_4 (t,x,y) &=& F_0 (t,x,y) + g(t) \psi_2 + F_4 (x,y) \psi_4.
 \eeqa
The functions $A_0,C_0,D_0$ and $F_0$ are again arbitrary and such that
they satisfy the equation (\ref{Schrot}) with $\Psi= (A_0,C_0,D_0,F_0)^t$. The other functions in (\ref{granphiuno}) are given by
 \beqa
 A_1 (x,y) &=& - i { e B \over 2} (\delta_4  x - \delta_3 y )  - i { \delta_2  \over 2} + ( \delta_7 + i \delta_5 ), \\
 C_2 (x,y) &=& - i { e B \over 2} (\delta_4  x - \delta_3 y )  + i
 { \delta_2  \over 2} + ( \delta_7 + i \delta_5 ), \\
 D_3 (x,y) &=& - i { e B \over 2} (\delta_4  x - \delta_3 y )  - i { \delta_2  \over 2} + ( \delta_8 + i \delta_6 ), \\
 F_4 (x,y) &=& - i { e B \over 2} (\delta_4  x - \delta_3 y )  + i
 { \delta_2  \over 2} + ( \delta_8 + i \delta_6 ) \label{efecuatro}
\eeqa and the functions $ f(t) $ and $ g(t) $ are obtained as \be
\label{fandg} f(t) = (\delta_9 -  i \delta_{11} ) e^{i
\wph_{\alpha\beta}  t }, \qquad g(t) = (\delta_{10} - i\delta_{12}
) e^{-i \wph_{\alpha\beta} t}. \ee with $\wph_{\alpha\beta}
=\wph(\alpha-\beta)={eB\over{2M}}(\alpha-\beta)$.

Let us here comment on this last solution. The prolongation method
has been applied to the set of equations
(\ref{systeme-jc}-\ref{systeme-jcdos}) and
(\ref{jctres}-\ref{jccuatro}) for arbitrary values of $B_1,\ B_2,\
B_3$ and $\ B_4$, and lead to the following sets of equations for
$f$ and $g$: \be \label{lafonefe} i { df \over dt} = {e \over 2M}
(B_3 - B_1) f  = {e \over 2M} (B_4 - B_2)f \ee and \be
\label{lafonge} i{dg \over dt} = {e \over 2M} (B_1 - B_3) g = {e
\over 2M} (B_2 - B_4) g. \ee They are always compatible in the
case under consideration, i.e. the one associated to the
Hamiltonian (\ref{hatotal}), and we get the explicit solution
(\ref{fandg}). In this context, a particular constant solution is
obtained when $\wph_{\alpha\beta}=0$ or equivalently when
$\alpha=\beta$. But if $ (B_3 - B_1) \ne (B_4 -B_2), $  the
equations (\ref{lafonefe}) and (\ref{lafonge}) admit the trivial
solution  $f(t)=g(t)=0$ which is a case that will be considered
later.

Let us insist on the fact that since we are here in the case where $f$ and $g$ are given by (\ref{fandg}), we will get
symmetries expressed by odd generators that will satisfy structure relations corresponding to a superalgebra.

The infinitesimal generators of the finite dimensional Lie algebra
of symmetries may be directly obtained from (\ref{jcchampvect})
with the preceding values (\ref{xi-jc}-\ref{granphicuatro}) but
once again to be able to get the finite transformations of
symmetries, we have to express the vector fields in terms of the
real variables $ u_\rho ,\nu_\rho,$ such that  $\psi_\rho= u_\rho
e^{i \nu_\rho }\ (\rho = 1,\ldots, 4)$. We thus get the following
basis of generators: \beqa X_1 &=&\partial_t -  { \wph_{\alpha
\beta} \over 2 } [( \partial_{\nu_3} - \partial_{\nu_1} ) + (
\partial_{\nu_4} -
\partial_{\nu_2} )], \\
X_2 &=& (x \partial_y - y \partial_x ) - {1 \over2} [ (\partial_{\nu_1} + \partial_{\nu_3} ) -
(\partial_{\nu_2} + \partial_{\nu_4} )],  \\
X_3 &=& \partial_x + {eB \over 2} \ y \sum_{\rho =1}^{4}  \partial_{\nu_\rho},  \\
X_4 &=& \partial_y -  {eB \over 2} \ x \sum_{\rho =1}^{4}  \partial_{\nu_\rho}, \\
X_5 &=&  \sum_{\rho =1}^{4}  \partial_{\nu_\rho},  \\
X_6 &=& \sum_{\rho =1}^{4}  u_\rho \partial_{u_\rho}, \\
\nonumber X_7 &=& \cos(\wph_{\alpha \beta}t - \nu_3 -\nu_ 1) u_3
\partial_{u_1} + {u_3 \over u_1}
\sin(\wph_{\alpha \beta}t - \nu_3 -\nu_ 1) \partial_{\nu_1}  \\
&+& \cos(\wph_{\alpha \beta}t - \nu_4 -\nu_ 2) u_4 \partial_{u_2}
+ {u_4 \over u_2} \sin(\wph_{\alpha \beta}t - \nu_3 -\nu_ 1)
\partial_{\nu_2} , \\\nonumber
X_8 &=& \cos(\wph_{\alpha \beta}t - \nu_3 -\nu_ 1) u_1
\partial_{u_3} -
{u_1 \over u_3} \sin(\wph_{\alpha \beta}t - \nu_3 -\nu_ 1) \partial_{\nu_3} \\
&+& \cos(\wph_{\alpha \beta}t - \nu_4 -\nu_ 2) u_2 \partial_{u_4}
-
{u_2 \over u_4} \sin(\wph_{\alpha \beta}t - \nu_3 -\nu_ 1)\partial_{\nu_4} ,\\
X_9 &=& (u_3 \partial_{u_3} - u_1 \partial_{u_1} ) + (u_4 \partial_{u_4}- u_2 \partial_{u_2} ), \\
\nonumber X_{10} &=& \sin(\wph_{\alpha \beta}t - \nu_3-\nu_ 1) u_3
\partial_{u_1} -  {u_3 \over u_1}
\cos(\wph_{\alpha \beta}t - \nu_3-\nu_ 1)\partial_{\nu_1} \\
&+& \sin(\wph_{\alpha \beta}t - \nu_4 -\nu_ 2) u_4 \partial_{u_2}
-
{u_4 \over u_2} \cos(\wph_{\alpha \beta}t - \nu_3 -\nu_ 1)\partial_{\nu_2} , \\
\nonumber X_{11}&=& - \sin(\wph_{\alpha \beta}t - \nu_3 -\nu_ 1) u_1
\partial_{u_3} - {u_1 \over u_3} \sin(\wph_{\alpha \beta}t - \nu_3
-\nu_ 1)
\partial_{\nu_3} \\
&-& \sin(\wph_{\alpha \beta}t - \nu_4 -\nu_ 2) u_2 \partial_{u_4}
-{u_2 \over u_4} \cos( \wph_{\alpha \beta}t - \nu_3 -\nu_ 1)
\partial_{\nu_4}, \\
X_{12} &=& (\partial_{\nu_3} - \partial_{\nu_1} ) +
(\partial_{\nu_4} -\partial_{\nu_2} ). \eeqa  The generators $X_j\ (
j=1, \ldots, 6)$ satisfy the same commutation relations than the
ones satisfied by the generators (\ref{algebra1}-\ref{algebra6}).
The other generators $X_j\ ( j=7, \ldots, 12)$ form an algebra
isomorphic to $su(2)^{\mathbb C}$ as in the Pauli case. The
corresponding Lie algebra of symmetries of (\ref{Schrot}) is thus
isomorphic to $ (so(2){ \dotplus } h(2)) \oplus su(2)^{\mathbb C}
\oplus \{X_1,X_6\}.$

\subsection{A Lie superalgebra of symmetries}
The integration of the preceding vector fields leads to the
corresponding finite transformations on the space-time coordinates
and wave functions. We get, for $X_1,$ the invariance under time
translation such that \be {\tilde \Psi} ( {\tilde t}, {\tilde x},
{\tilde y}) =\pmatrix{e^{{i\over 2} \wph_{\alpha \beta} \lambda_1}
\sigma_0  & 0 \cr 0 & e^{-{i\over 2} \wph_{\alpha \beta} \lambda_1
} \sigma_0 \cr} \Psi (\tilde t - \lambda_1, {\tilde x}, {\tilde y}
). \ee The integration of the vector field $X_2$ implies \be
{\tilde t } = t , \quad \quad {\tilde x} = x \cos \lambda_2 - y
\sin \lambda_2, \quad {\tilde y}  = x \sin \lambda_2 +  y \cos
\lambda_2 \ee and \be {\tilde \Psi} ( {\tilde t}, {\tilde x},
{\tilde y}) = \pmatrix{e^{-{i\over 2} \lambda_2} & 0  &  0 & 0\cr
0 &  e^{{i\over 2} \lambda_2} & 0 & 0 \cr 0& 0 & e^{-{i\over 2}
\lambda_2 } & 0 \cr 0 &0 & 0&  e^{{i\over 2} \lambda_2} \cr} \Psi
( {\tilde t},{\tilde x} \cos \lambda_2 + {\tilde y} \sin
\lambda_2,  - {\tilde x} \sin \lambda_2 + {\tilde y} \cos
\lambda_2 ). \ee The integration of the vector fields $X_3 $ and
$X_4$ implies the invariance under space translations such that
\be {\tilde \Psi} ( {\tilde t}, {\tilde x}, {\tilde y}) =
e^{{i\over 2} e B {\tilde y} \lambda_3 } \Psi (\tilde t , {\tilde
x} - \lambda_3 , {\tilde y} ) \ee and \be {\tilde \Psi} ( {\tilde
t}, {\tilde x}, {\tilde y}) = e^{- {i\over 2} e B {\tilde x}
\lambda_4 } \Psi (\tilde t , {\tilde x} , {\tilde y} - \lambda_4
). \ee The integration of the vector fields $X_5$ and $X_6$ lead
to phase and scale transformations of the wave functions \be
{\tilde \Psi} ( t , x ,   y ) = e^{i \lambda_5} \Psi ( t ,   x ,
y) \ee and \be {\tilde \Psi} ( t , x ,   y ) = e^{ \lambda_6} \Psi
( t , x , y). \ee The remaining vector fields $X_j\ (j=7,\ldots
12)$ lead to the transformations \beqa {\tilde \Psi} (t, x, y )
&=& \pmatrix{\sigma_0 & \lambda_7 e^{i \wph_{\alpha \beta}
t } \sigma_0 \cr   0 & \sigma_0 \cr} \Psi ( t , x , y), \\
{\tilde \Psi} ( t , x ,   y ) &=& \pmatrix{\sigma_0 & 0 \cr
\lambda_8  e^{- i \wph_{\alpha \beta} t } \sigma_0 & \sigma_0
\cr} \Psi ( t , x , y),\\
 {\tilde \Psi} ( t, x, y ) &=& \pmatrix{e^{- \lambda_9 } \sigma_0 &0 \cr 0 &
  e^{\lambda_9 } \sigma_0 \cr} \Psi ( t , x , y),\\
 {\tilde \Psi} ( t, x, y ) &=& \pmatrix{\sigma_0 & -i \lambda_{10}
e^{i \wph_{\alpha \beta} t } \sigma_0 \cr   0 & \sigma_0 \cr} \Psi
(
t , x , y), \\
  {\tilde \Psi} ( t , x ,   y ) &=& \pmatrix{\sigma_0 & 0
 \cr   -i \lambda_{11}  e^{- i \wph_{\alpha \beta} t } \sigma_0 & \sigma_0
\cr} \Psi ( t , x , y),
 \\
 {\tilde \Psi} ( t, x, y ) &=& \pmatrix{e^{- i \lambda_{12} } \sigma_0 &0 \cr 0 &
  e^{i \lambda_{12} } \sigma_0 \cr} \Psi ( t , x , y).
\eeqa

A specific matrix realization of the symmetry generators is
obtained from these transformations, when they are developped at
first order in the parameter $\lambda_i$. After some linear
combinations and redefinitions, we get the following generators
\beqa X_1 &=& \partial_t \ {\mathbb I} - i {\wph_{\alpha \beta }
\over 2}  \pmatrix{\sigma_0 & 0 \cr 0 &- \sigma_0\cr}, \  X_2 = (x
\partial_y - y \partial_x )\ {\mathbb I} + {
i\over 2} \pmatrix{\sigma_3 & 0 \cr 0 & \sigma_3 \cr}, \\
X_3 &=& \left( \partial_x - i {eB \over 2}\ y \right)  {\mathbb I},
\quad X_4 = \left( \partial_y +  i {eB \over 2} \ x \right)
 {\mathbb I},\quad
X_5 =- i X_6 =- i {\mathbb I} , \\
{\mathbb T}_+ (t) &=&X_7= - i X_{10}= e^{i\wph_{\alpha \beta }t
}\pmatrix{0  & \sigma_0 \cr 0 & 0 \cr}, \\
{\mathbb T}_- (t) &=&X_8= - i X_{11}=  e^{- i\wph_{\alpha
\beta }t }\pmatrix{0  & 0 \cr \sigma_0 & 0 \cr},\\
{\mathbb Y} &=& X_9=-i X_{12}  = \pmatrix{\sigma_0 & 0 \cr 0 & -
\sigma_0 \cr}, \eeqa where $ {\mathbb I} $ is the $4\times 4$
identity matrix. We see that to $X_1 ,$ we can associate the
generator \be \label{jc-hsym} {\mathbb H}= i\partial_t \ {\mathbb
I}+{\wph_{\alpha \beta } \over 2} \pmatrix{\sigma_0 & 0 \cr 0 & -
\sigma_0 \cr }, \ee which will be related later to a new
Hamiltonian refering to a symmetrized version of the JC
Hamiltonian. The generator $X_2 $ corresponds to the total angular
momentum \be {\mathbb J} = \pmatrix{J & 0 \cr 0 & J \cr } \ee and
$X_3 $ and $X_4 $ may be combined to give \be {\mathbb A}_-
=\pmatrix{A_- & 0 \cr 0 & A_- \cr},\quad {\mathbb A}_+
=\pmatrix{A_+ & 0 \cr 0 & A_+ \cr} , \ee where $A_-$ and $A_+$ are
given in (\ref{anni-jc}-\ref{crea-jc}).

The generators ${\mathbb T_+} (t),{\mathbb T_-} (t) $ and
${\mathbb Y}$ may now be associated to odd generators and,
together with ${\mathbb A_-},\ {\mathbb A_+}$ and ${\mathbb I}$,
form a $sh(2/2)$ superalgebra. As in the cases of the SUSY
harmonic oscillator and the Pauli systems, odd products may be
formed between $\{ {\mathbb A_-},\ {\mathbb A_+}\}$ and $\{
{\mathbb T_+} (t),\ {\mathbb T_-} (t) \}$ and among the possible
ones we get the following generators: \beqa \label{esemenos}
{\mathbb S}_- (t) &=& \sqrt{{\tilde \wph}} {\mathbb A}_+ {\mathbb
T}_- (t) = \sqrt{\wph} e^{-
i \wph_{\alpha \beta} t} \pmatrix{0 & 0 \cr A_+ & 0 \cr}, \\
\label{esemas}
 {\mathbb S}_+ (t) &=& \sqrt{{\tilde \wph}} {\mathbb A}_-
{\mathbb T}_+ (t) = \sqrt{{\tilde \wph}} e^{i \wph_{\alpha \beta}
t} \pmatrix{0 & A_- \cr 0 & 0 \cr}, \eeqa and \beqa\label{umenos}
{\mathbb U}_- (t) &=& \sqrt{{\tilde \wph}} {\mathbb A}_- {\mathbb
T}_- (t) = \sqrt{{\tilde \wph}} e^{- i \wph_{\alpha \beta} t}
\pmatrix{0 & 0 \cr A_- & 0 \cr}, \\ \label{umas}{\mathbb U}_+ (t)
&=& \sqrt{{\tilde \wph}} {\mathbb A}_+ {\mathbb T}_+ (t) =
\sqrt{{\tilde \wph}} e^{i \wph_{\alpha \beta} t} \pmatrix{0 & A_+
\cr 0 & 0 \cr}, \eeqa which satisfy the anticommutation relations
\be \label{eses} \{{\mathbb S}_- (t) , {\mathbb S}_+ (t)\} =
{\mathbb H}_0 + {{\tilde \wph} \over 2} {\mathbb Y} = {\tilde
\wph} \pmatrix{A_- A_+ &0 \cr 0 & A_+ A_- }
 \ee
and \be \label{us} \{{\mathbb U}_- (t) , {\mathbb U}_+ (t)\} =
{\mathbb H}_0 -  {{\tilde \wph} \over 2} {\mathbb Y} = {\tilde
\wph} \pmatrix{A_+ A_- &0 \cr 0 & A_- A_+ \cr}, \ee where we have
defined \be {\mathbb H}_0 = {\tilde \wph} \pmatrix{\left( A_+ A_-
+ {1 \over 2} \right) & 0 \cr 0 & \left( A_+ A_- + {1 \over 2}
\right)\cr}. \ee The two components of ${\mathbb H}_0$ are not
related to ${\mathbb H}_{JC}$ since they have only diagonal terms.
The non-zero commutation relations between the generators
${\mathbb S}_\pm (t) $ and ${\mathbb U}_\pm (t)$ are given by \be
\{ {\mathbb S}_- , {\mathbb U}_+ \} = - 2 i {\mathbb C}_+ =
{\tilde \wph}{({\mathbb A}_+ )}^2, \qquad \{ {\mathbb S}_+ ,
{\mathbb U}_- \} = - 2 i {\mathbb C}_- = {\tilde \wph} {({\mathbb
A}_- )}^2. \ee
\begin{table}
\begin{center}
\begin{tabular}{|c|c|c|c|c||c|c|c|c|}\cline{2-9} \multicolumn{1}{c|}{} & ${\mathbb
H}_0 $ & ${\mathbb C}_- $ & ${\mathbb C}_+ $ & ${\mathbb Y} $& ${\mathbb S}_- $ & ${\mathbb S}_+ $ & ${\mathbb U}_- $ & ${\mathbb U}_+ $ \\
\cline{1-9} ${\mathbb H}_0 $  & $0$  & $ - 2 {\tilde \wph}{\mathbb
C}_- $ & $2 {\tilde \wph}{\mathbb C}_+ $ & $ 0 $ & ${\tilde \wph}
{\mathbb S}_- $ & $-{\tilde \wph} {\mathbb S}_+  $
& $- {\tilde \wph} {\mathbb U}_- $ & ${\tilde \wph} {\mathbb U}_+ $ \\
\hline ${\mathbb C}_- $  & $2 {\tilde \wph}{\mathbb C}_- $ & $ 0$
& $- {\tilde \wph} {\mathbb H}_0 $ & $0$ & $i {\tilde
\wph}{\mathbb U}_- $
& $0$ & $0 $ & $i {\tilde \wph}{\mathbb S}_+$ \\
\hline${\mathbb C}_+ $  & $- 2{ \tilde \wph}{\mathbb C}_+ $ &
${\tilde \wph} {\mathbb H}_0  $ & $0$ & $0$ & $0$ & $-i{\tilde
\wph}{\mathbb U}_+  $   & $- i {\tilde \wph}{\mathbb S}_- $ &
 $0$ \\
\hline ${\mathbb Y}$  & $0$  & $ 0 $ & $0 $ & $0$ & $-2{\mathbb
S}_-  $ & $2 {\mathbb S}_+ $ & $-2{\mathbb U}_-$ & $2{\mathbb U}_+
$
\\ \hline \hline ${\mathbb S}_- $ & $-{\tilde \wph} {\mathbb S}_- $ & $ -i {\tilde \wph}{\mathbb
U}_- $ & $0$ & $2 {\mathbb S}_- $ & $0 $ & $ {\mathbb H}_0 + {\tilde \wph} {\mathbb Y} / 2 $ & $0 $ & $ - 2i {\mathbb C}_+ $  \\
\hline
${\mathbb S}_+ $  & ${\tilde \wph} {\mathbb S}_+ $ & $ 0$ & $i{\tilde \wph}{\mathbb U}_+ $ & $-2 {\mathbb S}_+ $ & ${\mathbb H}_0 + {\tilde \wph} {\mathbb Y} / 2$ & $0$ & $-2i {\mathbb C}_- $ & $0$\\
\hline
${\mathbb U}_- $ & ${\tilde \wph} {\mathbb U}_- $ & $ 0$ & $i {\tilde \wph} {\mathbb S}_- $ & $2 {\mathbb U}_- $ & $0 $ & $- 2i{\mathbb C}_- $ & $0 $ & $ {\mathbb H}_0 - {\tilde \wph}{\mathbb Y} /2  $ \\
\hline
${\mathbb U}_+ $  & $- {\tilde \wph} {\mathbb U}_+$ & $ - i {\tilde \wph} {\mathbb S}_+$ & $0$ & $ -2 {\mathbb U}_+ $ & $-2i {\mathbb C}_+$ & $0$ & ${\mathbb H}_0 - {\tilde \wph}{\mathbb Y} /2 $ & $0$ \\
\hline
\end{tabular}
\end{center} \caption{Super-commutation relations of an $ osp(2/2)$
superalgebra.} \label{tab:super-jc}
\end{table}
Now the set $\{\mathbb H, {\mathbb H}_0 , {\mathbb C}_\pm, \mathbb
Y, {\mathbb S}_\pm (t) , {\mathbb U}_\pm (t),  \mathbb J, {\mathbb
A}_\pm, \mathbb I, {\mathbb T}_\pm (t) \}  $ closes a superalgebra.
We see that $\mathbb H$ commutes with all the generators. The
commutation relations between the generators ${\mathbb H}_0 ,
{\mathbb C}_\pm, \mathbb Y, {\mathbb S}_\pm (t)$ and  ${\mathbb
U}_\pm(t) $ are given in table \ref{tab:super-jc}. These last
generators form a superalgebra isomorphic to $osp(2/2). $ The
non-zero super-commutation relations between the generators $
{\mathbb J}, {\mathbb A}_\pm, \mathbb I,{\mathbb T}_\pm (t),$ are
now given by \be [{\mathbb J} , {\mathbb A}_+ ] = {\mathbb A}_+ ,
\qquad [{\mathbb J} , {\mathbb A}_- ] = - {\mathbb A}_- \ee and \be
[{\mathbb A}_- , {\mathbb A}_+ ] = {\mathbb I} = \{ {\mathbb T}_-
(t) , {\mathbb T}_+ (t) \}, \ee leading to the superalgebra $so(2){
\dotplus } sh(2/2)$. Finally the super-commutation relations between
the two sets are presented in table \ref{tab:dos-ensembes}. So we
find a structure isomorphic to the superalgebra $( so(2) { \dotplus
} osp(2/2) ) { \dotplus } sh(2/2).$ Let us insist on the fact that
the existence of such a superalgebra does not implies the presence
of supersymmetries for the original Hamiltonian (\ref{hatotal}).
Indeed, no $Q$-type supercharges may be constructed from the
preceding symmetries. In the last subsections,  SUSY JC models will
be constructed and the symmetries and supersymmetries will be given.

\begin{table}
\begin{center}
\begin{tabular}{|c|c|c|c|c||c|c|c|c|}\cline{2-9} \multicolumn{1}{c|}{} & ${\mathbb
H}_0 $ & ${\mathbb C}_- $ & ${\mathbb C}_+ $ & ${\mathbb Y} $ & ${\mathbb S}_- $ & ${\mathbb S}_+ $ & ${\mathbb U}_- $ & ${\mathbb U}_+ $ \\
\cline{1-9} ${\mathbb J} $  & $0$  & $-2{\mathbb C}_-$  & $
2{\mathbb C}_+ $ & $0$ & $ {\mathbb S}_- $ & $ - {\mathbb S}_+ $ &
$- {\mathbb U}_-  $
& ${\mathbb U}_+ $    \\
\hline ${\mathbb A}_- $  & ${\tilde \wph}{\mathbb A}_- $ & $ 0$ &
$i {\tilde \wph} {\mathbb A}_+ $ & $0$ & $\sqrt{\tilde
\wph}{\mathbb T}_- $
& $0$ & $0 $ & $\sqrt{\tilde \wph}{\mathbb T}_+ $ \\
\hline${\mathbb A}_+ $  & $- { \tilde \wph}{\mathbb A}_+ $ & $ -i
{\tilde \wph} {\mathbb A}_-  $ & $0$ & $0$ & $0$ & $-\sqrt{\tilde
\wph}{\mathbb T}_+ $   & $- \sqrt{\tilde \wph}{\mathbb T}_- $ &
 $0$ \\
\hline ${\mathbb I}$  & $0$  & $ 0 $ & $0 $ & $0$ & $0$ & $0$ &
$0$ & $0$
\\ \hline \hline  ${\mathbb T}_-$ & $0$ & $0$ & $0$ & $2 {\mathbb T}_- $ & $0 $ & $\sqrt{\tilde \wph} {\mathbb A}_- $ & $0 $
& $ \sqrt{\tilde \wph } {\mathbb A}_+ $  \\
\hline
${\mathbb T}_+ $  & $0$ & $ 0$ & $0 $ & $-2 {\mathbb T}_+ $ & $\sqrt{\tilde \wph} {\mathbb A}_+ $ & $0$ & $\sqrt{\tilde \wph} {\mathbb A}_- $ & $0$\\
\hline
\end{tabular}
\end{center} \caption{Commutation relations between $( so(2) { \dotplus
} osp(2/2))$ and  $sh(2/2).$} \label{tab:dos-ensembes}
\end{table}

\subsection{A supersymmetric JC model}

The generator (\ref{jc-hsym}) when acting on the space of
solutions of (\ref{Schrot}), corresponds to a symmetrized version
of the Hamiltonian (\ref{hatotal}) given by \be {\mathbb H}= H_T +
{\wph_{\alpha \beta}\over 2} {\mathbb Y}= \pmatrix{H_{JC} -
{eB\over 4M} (\alpha+\beta) \sigma_0 & 0 \cr 0& H_{JC} - {eB\over
4M} (\alpha+\beta) \sigma_0 \cr }. \ee From the preceding results,
it is easy to show that the symmetries of this new Hamiltonian are
given by the set $\{{\mathbb H } , {\mathbb H}_0 , {\mathbb
C}_\pm, \mathbb Y, \mathbb J, {\mathbb A}_\pm, \mathbb I, {\mathbb
T}_\pm (0) \}$, all of these generators being time independent.
Indeed, ${\mathbb H}$ can be seen as a particular $H_T$ as given
in (\ref{hatotal}) where $\alpha=\beta$ and thus $\wph_{\alpha
\beta}=0$.

It admits also the supersymmetries ${\mathbb S}_\pm(0)$ and ${\mathbb U}_\pm(0)$ which
are now time independent and satisfy again ($\ref{eses}$) and ($\ref{us}$). None of them are the supercharges
of ${\mathbb H}$.

Let us now show that, for a specific value of
$\alpha+\beta$, the symmetrized Hamiltonian ${\mathbb H}$ can be made supersymmetric.
Indeed, if we take $ (\alpha +\beta) = - {e E \over
{8 M^2 B}}$, we can define \cite{kn:AnLe}

\be {\mathbb Q}_+ = {{\bar \kappa} \over 2 \sqrt{{\tilde \wph}}}
{\mathbb T}_+ (0) + \sqrt{{\tilde \wph}} \pmatrix{0& ({Q}_+ - {Q}_-
)\cr 0&0\cr} \ee and \be {\mathbb Q}_- = {\kappa \over 2
\sqrt{{\tilde \wph}}} {\mathbb T}_- (0) + \sqrt{{\tilde \wph}}
\pmatrix{0& 0\cr ({Q}_- - {Q}_+ )&0\cr}. \ee We thus have \be
\{{\mathbb Q}_+ , {\mathbb Q}_- \} = {\mathbb H}, \qquad [{\mathbb
H}, {\mathbb Q}_\pm ] =0. \ee The time independent generators
${\mathbb H}, {\mathbb Y}, {\mathbb J}, {\mathbb A}_\pm , {\mathbb
I}$ and ${\mathbb Q}_\pm $  form a superalgebra of supersymmetries
of ${\mathbb H}.$ The additionnal super-commutation relations are
\be [{\mathbb Y} , {\mathbb Q}_\pm ] = \pm 2 {\mathbb Q}_\pm. \ee If
we include the generators ${\mathbb T}_\pm (0)$ as symmetries of
${\mathbb H}$ , we get the following superalgebra $\{{\mathbb H},
{\mathbb Y}, {\mathbb J}, {\mathbb A}_\pm , {\mathbb I}, {\mathbb
Q}_\pm , {\mathbb Q}_0, {\mathbb T}_\pm (0)\},$ where \be {\mathbb
Q}_0 = \pmatrix{({ Q}_+ - {Q}_- )& 0 \cr 0 &({Q}_+ - {Q}_- ) \cr}.
\ee Indeed we have \be \{ {\mathbb T}_+ (0) , {\mathbb Q}_-  \} =
{\kappa \over 2 \sqrt{{\tilde \wph}}} {\mathbb I} - \sqrt{{\tilde
\wph}} {\mathbb Q}_0, \qquad \{ {\mathbb T}_- (0), {\mathbb Q}_+  \}
= {{\bar \kappa} \over 2 \sqrt{{\tilde \wph}}} {\mathbb I} +
\sqrt{{\tilde \wph}} {\mathbb Q}_0. \ee This superalgebra may be
written as $ (\{ {\mathbb H}, {\mathbb Y}, {\mathbb Q}_\pm \} \oplus
\{{\mathbb J} \} )  { \dotplus } \{ {\mathbb A}_\pm , {\mathbb I},
{\mathbb T}_\pm (0) , {\mathbb Q}_0 \}. $

In the approach of Andreev and Lerner \cite{kn:AnLe}, the
preceding Hamiltonian  ${\mathbb H}$ has been generalized to
${\mathbb H} (\varphi ) $ where $\varphi$ is an arbitrary phase.
Indeed,  ${\mathbb H} (\varphi )$ is block diagonal where, up to
the addition of a multiple of the identity, the first block is the
JC Hamiltonian (\ref{hajainescu}) and the second one is obtained
from it by changing $ {\cal A} \mapsto e^{- i \varphi} {\cal A} $
and $ {\cal A}^\dagger \mapsto e^{i \varphi} {\cal A}^\dagger. $
With respect to our approach, it is associated to the original set
of equations (\ref{systeme-jc}-\ref{systeme-jcdos}) and the new
set (\ref{jctres}-\ref{jccuatro}) where $ E_3 = - {\tilde E}_3 = E
e^{i \varphi} $  and $ E_4 =  {\tilde E}_4 = E e^{-i \varphi}. $
The algebra of symmetries is the same as for the case $\varphi=0$
studied before, so all the results about the existence of
supersymmetry transformations remain valid. The only changes are
in the following generators \be {\mathbb T}_+ (\varphi , t) =
e^{i\wph_{\alpha \beta }t }\pmatrix{0 & \pmatrix{1&0 \cr 0 & e^{i
\varphi } \cr} \cr 0 & 0 \cr},
 {\mathbb T}_- (\varphi, t) = e^{-
i\wph_{\alpha \beta }t } \pmatrix{0 & 0 \cr \pmatrix{1&0 \cr 0 &
e^{- i \varphi } \cr} & 0 \cr}. \ee It follows that the generators
$ {\mathbb S}_\mp (t )$  and $ {\mathbb U}_\mp (t )$ given
(\ref{esemenos}-\ref{esemas}) and (\ref{umenos}-\ref{umas}) in are
now written as $ {\mathbb S}_\mp (\varphi , t ) = \sqrt{\tilde
\wph } {\mathbb A}_\pm {\mathbb T}_\mp (\varphi ,t) $ and
${\mathbb U}_\mp (\varphi , t ) = \sqrt{\tilde \wph } {\mathbb
A}_\mp{\mathbb T}_\mp (\varphi , t ) . $

The supercharges are found to be \beqa {\mathbb Q}_+ (\varphi )
&=& {{\bar \kappa} \over 2 \sqrt{{\tilde \wph}}} {\mathbb T}_+
(\varphi, 0) + \sqrt{{\tilde \wph}} \pmatrix{0& ( e^{i\varphi}
{Q}_+ - {Q}_-
)\cr 0&0\cr} ,\\
{\mathbb Q}_- (\varphi ) &=& {\kappa \over 2 \sqrt{{\tilde \wph}}}
{\mathbb T}_- (\varphi ,  0) + \sqrt{{\tilde \wph}} \pmatrix{0&
0\cr ( e^{-i \varphi } {Q}_- - {Q}_+ )&0\cr} \eeqa and satisfy \be
\{{\mathbb Q}_+ (\varphi) , {\mathbb Q}_- (\varphi )\} = {\mathbb
H} (\varphi ), \qquad [{\mathbb H} (\varphi ), {\mathbb Q}_\pm
(\varphi )] =0. \ee The last generator that is modified is \be
{\mathbb Q}_0 (\varphi) = \pmatrix{({ Q}_+ - {Q}_- )& 0 \cr 0 &(
e^{i \varphi} { Q}_+ - e^{- i \varphi}{ Q}_- ) \cr}.\ee

\subsection{The usual supersymmetric structure}

Another SUSY version of the JC model may be deduced from our
preceding considerations. Let us refer it as the standard or
strong coupling limit one in reference with the literature
\cite{kn:BHN,kn:OrSa,kn:NaVh}. If we start again with the system
of equations (\ref{systeme-jc}-\ref{systeme-jcdos}) and
(\ref{jctres}-\ref{jccuatro}) for which the symmetries have been
determined for arbitrary values of the parameters $B_1,\ B_2, \
B_3, \ B_4,$ we can in particular take $B_1 =- B_2 = B,$ while $
B_3 = -3B$ and $B_4 = - B.$ This is the case where the equations
(\ref{lafonefe}) and (\ref{lafonge}) admit the trivial solution
$f(t)=g(t)=0$ and such that no symmetries associated with odd
generators appear. This means that, by the prolongation method, it
will be impossible to get a superalgebra of symmetries.

Meanwhile, if we choose $E_2 = - E_1 = E_4 = -E_3 = {\tilde E}_3 =
{\tilde E}_4 = E = M / \sqrt{2eB},$ it is possible to write the
evolution equations (\ref{systeme-jc}-\ref{systeme-jcdos}) and
(\ref{jctres}-\ref{jccuatro}) as  \be (i \partial_t -
{\mathbb H}_{JC}) \Psi(t,x,y) =0, \ee where ${\mathbb H }_{JC} $ has
the standard SUSY form\cite{kn:FeHu}
\be \label{hjc-usual}
 {\mathbb H}_{JC} =
{\tilde \wph} \pmatrix{ { \tilde {\cal A} }^\dagger {\tilde {\cal
A} } & 0 \cr 0 & { \tilde {\cal A} } { \tilde {\cal A} }^\dagger
\cr}, \ee with \be
 {\tilde {\cal A} } = {\cal A } + i \sigma_+ , \qquad  { \tilde {\cal A} }^\dagger =
 {\cal A }^\dagger - i \sigma_- .
\ee These last operators satisfy \be [{\tilde {\cal A} }, { \tilde
{\cal A} }^\dagger ] = \sigma_0 + \sigma_3. \ee Note that the two
components of the Hamiltonian ${\mathbb H }_{JC} $ are closely
related to the Hamiltonian (\ref{hajainescu}). Indeed, we have \be
\label{standjcha} {\mathbb H }_{JC} = \pmatrix{ H_{JC}  & 0 \cr 0
& H_{JC} + {\tilde \wph} \pmatrix{0 & 0 \cr 0 & 1 \cr } \cr}, \ee
when $\kappa = i {\tilde \wph } $.  The standard supercharges are
given by \be { \tilde {\cal Q} }_+ = { \tilde \wph }^{1/2}
\pmatrix{0& { \tilde {\cal A} }^\dagger \cr 0 & 0 \cr}, \qquad {
\tilde {\cal Q } }_- = { \tilde \wph }^{1/2}\pmatrix{0 & 0 \cr
{\tilde {\cal A} } & 0 \cr}, \ee satisfying the following
relations \be { \mathbb H }_{JC} = \{ { \tilde {\cal Q} }_+ , {
\tilde {\cal Q} }_- \}, \qquad {({ \tilde {\cal Q} }_\pm )}^2 = 0
,\qquad [ {\mathbb H }_{JC} , { \tilde {\cal Q} }_\pm ] =0. \ee
Again these supercharges can not be obtained from the product of
symmetries determined by the prolongation method.

Let us finally
mention that such a Hamiltonian is a particular case of a matrix
SUSY one where the quantities $ \tilde {\cal A} $ and ${\tilde
{\cal A} }^\dagger $ would correspond to elements of the algebra  $ h(2)
\oplus su(2) , $ that is linear combinations of the generators of
this algebra. Different assumptions may thus be imposed on the
commutator $[ {\cal A } , {{\cal A }}^\dagger ] $ \cite{kn:NaVh}.
In the canonical case, that is the case where the commutator is a
multiple of the identity, the prolongation method reproduce all
the dynamical supersymmetries. This was the case for the SUSY
harmonic oscillator and the Pauli Hamiltonians. In the
non-canonical case, like for example the JC model, the
supercharges are not obtained from the prolongation method but may
be constructed by the standard structure of the Hamiltonian like
in (\ref{hjc-usual}) for the JC model.
\section{Conclusion}
We have shown that the prolongation method used for
finding symmetries of classical as well as quantum mechanical
systems may be useful to determine the supersymmetries of SUSY quantum
mechanical systems. We took simple examples, like the SUSY
harmonic oscillator and the Pauli equations to improve the method.
Indeed, we already knew the kind of kinematical and dynamical
superalgebras we were searching for. This was very helpful to be
able to get new results on the JC model. First we determine the
Lie algebra of symmetries for the usual 2 by 2 matrix model.
Second, we gave the symmetry superalgebra for a generalized
version which is a amplification of the usual JC model to a 4 by 4
matrix representation. Finally, two ways of getting SUSY versions
where given. In the first case the supersymmetry was present only
if we admit a specific shifting in the JC Hamiltonian. In the
second case, the supersymmetry appeared due to the fact that the
amplification of the JC model is similar to the one of the SUSY
harmonic oscillator but to do it was necessary to take the
coupling constant between the electromagnetic field and the atom
as a linear function of the frequency of these fields. In all
these cases, the detuning between the electromagnetic field and
atom frequencies has been assumed to be equal to zero.
\section*{Acnowledgments} Nibaldo Alvarez dedicates this article
to the memory of his friend  Cristian Gar\'{\i}n Reyes, {\it el
artista poeta chileno decanta  pasos en un bosque transitorio con
sus tritones torcidos de color}. The authors' research was
partially supported by research grants from NSERC of Canada and
FQRNT of Qu\'ebec. N.~A.~M acknowledges financial support from the
ISM. \nolinebreak

\end{document}